\documentclass[showkeys,superscriptaddress,floatfix,prd,10pt,aps]{revtex4-2}
\usepackage{graphicx,epstopdf}
\usepackage[caption=false]{subfig}
\usepackage{appendix}
\usepackage[T1]{fontenc}
\usepackage{lmodern}
\usepackage[dvipsnames,x11names]{xcolor}
\usepackage[colorlinks=true,linkcolor=NavyBlue,citecolor=ForestGreen,urlcolor=NavyBlue]{hyperref}
\usepackage[sort&compress]{natbib}
\usepackage{lipsum}
\usepackage{morefloats}
\usepackage[pdf]{pstricks}
\usepackage{amsmath}
\usepackage{amssymb}
\usepackage{amsfonts}
\usepackage{rotating}
\usepackage{cancel}
\usepackage{mathtools}
\usepackage{bbm}
\usepackage{dsfont}
\usepackage{bbold}
\usepackage{multirow}
\usepackage{ulem}
\usepackage{physics}
\usepackage{orcidlink}
\usepackage{colortbl}

\def\pslash{p\!\!\!\slash }
\def\qslash{q\!\!\!\slash }
\def\xslash{x\!\!\!\slash }
\def\eslash{\varepsilon\!\!\!\slash }

\def\vel{\left|}
\def\ver{\right|}

\begin{document}

\title{Analysis of the isospin eigenstate $\bar D \Sigma_c$, $\bar D^{*} \Sigma_c$, and $\bar D \Sigma_c^{*}$ pentaquarks by their electromagnetic properties}

\author{Ula\c{s} \"{O}zdem\orcidlink{0000-0002-1907-2894}}%
\email[]{ulasozdem@aydin.edu.tr}
\affiliation{Health Services Vocational School of Higher Education, Istanbul Aydin University, Sefakoy-Kucukcekmece, 34295 Istanbul, T\"{u}rkiye}

 
\begin{abstract}
To shed light on the nature of the controversial and not yet fully understood exotic states, we are carrying out a systematic study of their electromagnetic properties. The magnetic moment of a hadron state is as fundamental a dynamical quantity as its mass and contains valuable information on the deep underlying structure.  In this study, we use the QCD light-cone sum rule to extract the magnetic moments of the $\mathrm{P_{c}(4312)}$, $\mathrm{P_{c}(4380)}$, and $\mathrm{P_{c}(4440)}$ pentaquarks by considering them as the molecular picture with spin-parity $\mathrm{J^P= \frac{1}{2}^-}$, $\mathrm{J^P= \frac{3}{2}^-}$, and $\mathrm{J^P= \frac{3}{2}^-}$, respectively. We define the isospin of the interpolating currents of these states, which is the key to solving the puzzle of the hidden-charm pentaquark states, to make these analyses more precise and reliable. We have compared our results with other theoretical predictions that could be a useful complementary tool for the interpretation of the hidden-charm pentaquark sector, and we observe that they are not in mutual agreement with each other. We have also calculated higher multipole moments for spin-3/2 $\bar D^{*}  \Sigma_c$ and $\bar D \Sigma_c^{*}$ pentaquarks, indicating a non-spherical charge distribution.
\end{abstract}

\maketitle

\section{Introduction}\label{motivation}
In recent years, significant experimental progress has been made in the study of exotic hadrons, in particular in the observation of pentaquark states with hidden-charms, which has led to a great deal of interest in the study of pentaquarks. In 2015, two pentaquark states, named $\mathrm{P_{c}(4380)}$ and $\mathrm{P_{c}(4450)}$, were discovered by the LHCb Collaboration in the $J/\psi\,p$ invariant mass distribution of the $\Lambda_b \rightarrow J/\psi\,p\,K$ decay~\cite{Aaij:2015tga}. Four years later, with the inclusion of additional data, it was discovered that the $\mathrm{P_{c}(4450)}$ pentaquark state is composed of two narrow overlapping peaks, $\mathrm{P_{c}(4440)}$, and $\mathrm{P_{c}(4457)}$. Furthermore, a new narrow state, the $\mathrm{P_{c}(4312)}$, has been reported~\cite{Aaij:2019vzc}. Note that the $\mathrm{\mathrm{P_c}(4380)}$ pentaquark state could neither be confirmed nor refuted in the updated analysis. In 2020 the LHCb Collaboration announced the first observation of the hidden-charm pentaquark state with strangeness, named $\mathrm{P_{cs}(4459)}$~\cite{LHCb:2021chn}. The LHCb Collaboration announced recently also the discovery of two new narrow hidden-charm strange pentaquark states, $\mathrm{P_{cs}(4338)}$, with minimal quark content ($c\bar c uds$)~\cite{Collaboration:2022boa}. 
The measured masses and widths of the four $c\bar c uud$ pentaquark states are
 %
\begin{align}
\notag  \rm{P_c(4312)}:M&=4311.9 \pm 0.7^{+6.8}_{-0.6}\,\rm{MeV}\, ,
\,\,\,
\Gamma=9.8\pm2.7^{+3.7}_{-4.5}\,\rm{MeV}\,,\nonumber\\
 \rm{P_c(4380)}:M&=4380 \pm 8.0 \pm29\,\rm{MeV}\, ,
 \,\,\,\,\,
\Gamma=205\pm18\pm86\,\rm{MeV}\,,\nonumber\\
\rm{ P_c(4440)}:M&=4440.3 \pm 1.3^{+4.1}_{-4.7}\,\rm{MeV}\,, 
\,\,\,
\Gamma=20.6\pm4.9^{+8.7}_{-10.1}\,\rm{MeV}\,,\nonumber\\
\rm{P_c(4457)}:M&=4457.3\pm0.6^{+4.1}_{-1.7}\,\rm{MeV}\,,
 \,\,\,
\Gamma=6.4\pm2.0^{+5.7}_{-1.9}\,\rm{MeV}\,.
\end{align}


The discovery of the aforementioned pentaquarks immediately stimulated discussion of their properties and internal structure.  Systematic study of their internal structure and nature can be a major step forward in our understanding of the non-perturbative behavior of the strong interaction. Hence, there have been a large number of theoretical
studies trying to determine their masses, radiative decays, magnetic moments, and their production rates in the $\Lambda_b$ decay or the inclusive processes. Further discussions on the progress of the observed and possible pentaquarks and the other exotic states can be found in Refs.~\cite{Esposito:2014rxa,Esposito:2016noz,Olsen:2017bmm,Lebed:2016hpi,Nielsen:2009uh,Brambilla:2019esw,Agaev:2020zad,Chen:2016qju,Ali:2017jda,Guo:2017jvc,Liu:2019zoy,Yang:2020atz,Dong:2021juy,Dong:2021bvy,Meng:2022ozq, Chen:2022asf}, both in theory and experiment. However, despite all efforts, the true nature of these states remains one of the most important and challenging problems in hadron physics.
As a result, further theoretical and experimental studies are needed to shed light on the nature of these and other exotic states.

The electromagnetic properties of the pentaquark states have not received much attention, though the mass spectra, production mechanisms, and decay behavior of the pentaquark states have received a great deal of attention from both theoretical and experimental sides~\cite{Wang:2016dzu, Ozdem:2018qeh, Ortiz-Pacheco:2018ccl,Xu:2020flp,Ozdem:2021btf, Li:2021ryu,Ozdem:2021ugy,Gao:2021hmv, Ozdem:2022iqk, Ozdem:2022vip, Ozdem:2022kei, Wang:2022tib,Ozdem:2023htj,Wang:2023iox,Wang:2023ael,Guo:2023fih}. The magnetic moment is another observable parameter that may provide important information about the quark-gluon structure of hadrons and their fundamental nature and underlying dynamics. 
The magnetic moments of pentaquarks are a measure of their ability to interact with magnetic fields, and they are a prominent property for understanding the behavior of these states. In the electro- or photo-production of pentaquarks, different magnetic moments have an effect on both the differential and total cross-sections. 
Therefore, it is clear that a determination of the magnetic moments of the pentaquark states is extremely important to verify their nature. Motivated by this, in this study, we investigate the magnetic moments of the $\mathrm{P_{c}(4312)}$, $\mathrm{P_{c}(4380)}$, and $\mathrm{P_{c}(4440)}$ pentaquarks, assuming that these pentaquark states are $\bar D \Sigma_c$, $\bar D^{*} \Sigma_c$, and $\bar D \Sigma_c^{*}$ molecular states with spin-parity $\mathrm{J^P= \frac{1}{2}^-}$, $\mathrm{J^P= \frac{3}{2}^-}$, and $\mathrm{J^P= \frac{3}{2}^-}$, respectively. To make these analyses more precise and reliable, we define the isospin of the interpolating currents of these states, which is the key to solving the puzzle of the $\mathrm{P_c}$ states. Thanks to this, these interpolating currents couple potentially to the color singlet-singlet type $\mathrm{P_c}$ states rather than to the meson-baryon scattering states or thresholds. We should mention here that it is possible to study the magnetic moment of the $\mathrm{P_{c}(4457)}$ state as a $\bar D^* \Sigma_c^*$ molecular state. However, since the $\bar D^* \Sigma_c^*$ state is a spin-5/2 state, it can cause some additional complications in the calculations, which can affect the reliability of the analysis. Therefore, the magnetic moment of the $\bar D^* \Sigma_c^*$ state presents additional challenges and is beyond the scope of the present study. The magnetic moments belong to the non-perturbative regime of QCD and hence, a reliable non-perturbative method is needed to evaluate these physical observables. The QCD light-cone sum rule is one of the most efficient tools for the calculation of non-perturbative parameters~\cite{Chernyak:1990ag, Braun:1988qv, Balitsky:1989ry}.

The present work is organized as follows.  In Sec. \ref{formalism}, we obtain the QCD light-cone sum rules for isospin eigenstate $\bar D \Sigma_c$, $\bar D^{*} \Sigma_c$, and $\bar D \Sigma_c^{*}$ pentaquarks. The numerical analysis together with the discussion is presented in Sec. \ref{numerical}.  The last section is reserved for the summary. In the Appendix, we list the explicit expressions obtained for the magnetic moment of the $\bar D \Sigma_c$ pentaquark state with isospin-1/2.

\begin{widetext}
 
\section{Electromagnetic properties of the isospin eigenstate $\bar D \Sigma_c$, $\bar D^{*} \Sigma_c$, and $\bar D \Sigma_c^{*}$ pentaquarks within the QCD light-cone sum rules}\label{formalism}

The initial attempt to study the magnetic moments of the spin-1/2 and spin-3/2 pentaquark states (hereafter referred to as $\mathrm{P_c}$ and $\mathrm{P_c^*}$, respectively) by QCD light-cone sum rules is to introduce the following correlation functions,
 \begin{align} \label{edmn01}
\Pi(p,q)&=i\int d^4x e^{ip \cdot x} \langle0|T\left\{J^{\mathrm{P_c}}(x)\bar{J}^{\mathrm{P_c}}(0)\right\}|0\rangle _\gamma \, , \\
\Pi_{\mu\nu}(p,q)&=i\int d^4x e^{ip \cdot x} \langle0|T\left\{J_\mu^{\mathrm{P_c^*}}(x)\bar{J}_\nu^{\mathrm{P_c^*}}(0)\right\}|0\rangle _\gamma \,, \label{Pc101}
\end{align}
where $T$ is the time ordered product, the $\gamma$ is the weak external electromagnetic field. The $J^{\mathrm{P_c}}(x)$ and $\mathrm{J^{P^*_{c}}}(x)$ are the interpolating currents for the spin-$ 1/2^-$ and spin-$ 3/2^-$ states, respectively. 

It is important to note that a number of potential interpolating currents can be formulated for hidden-charm pentaquark states. However, the number of possible interpolating currents that can be written can be slightly reduced when taking into account the QCD sum rules and the quantum numbers of the states under study, such as spin, parity, and isospin. In order to comprehend the structure and properties of a hadron, it is possible to employ an effective approach which involves assigning a structure that is suitable for the task and subsequently calculating the relevant properties. The theoretical findings that result from the aforementioned calculations can significantly aid in the comprehension of the nature and substructure of the hadron. To achieve this objective, it is imperative to identify suitable interpolating currents, comprised of quark fields that align with the valence quark content and quantum numbers of hidden-charm pentaquark states. The formation of diquarks in color antitriplet is corroborated by the attractive interaction that is induced by the one-gluon exchange. It can be reasonably inferred from QCD sum-rules that the scalar $(C\gamma_5)$ and axial-vector $(C\gamma_\mu)$ diquark states are the most plausible configurations \cite{Wang:2010sh,Kleiv:2013dta}. The present study considers the molecular hidden charm pentaquark states $\bar D \Sigma_c$, $\bar D^{*} \Sigma_c$, and $\bar D \Sigma_c^{*}$.  The ground state heavy baryons are made from the heavy quark Q with spin-parity $J^P = 1/2^+$ and a light diquark system with spin-parity $ J^P = 0^+$ ($\Lambda$-type)/ and $1^+$ ($\Sigma$-type) moving in a s-wave state relative to the heavy quark \cite{Korner:1994nh}. For the qq diquark states with $q = u, d ~\mbox{or}~ s$, we have to resort to the $C\gamma_\mu$ diquark states to satisfy the Fermi-Dirac statistic. In short, for the  $\Sigma_c^{(*)}$ baryon we prefer the $C\gamma_\mu$ diquark with different structure of  c-quark $(\gamma^\alpha \gamma_5 c^c~ \mbox{or} ~c^c$) in the present work. Then to obtain molecular hidden-charm tetraquarks, we combine this structures with $\bar D ~(J^P = 0^-)$ and $\bar D^{*} ~(J^P = 1^-)$ mesons.  As a result, the following interpolating currents have been written for the pentaquark states analyzed in this study, taking into account all the possibilities mentioned above: 
%
%
\begin{align}\label{curpcs2}
J^{\bar D \Sigma_c}(x)& =\frac{1}{\sqrt{3}}\mid \bar D^0(x) \Sigma_c^+(x) \rangle \, \mp \sqrt{\frac{2}{3}}\mid \bar D^-(x) \Sigma_c^{++}(x) \rangle  
= \frac{1}{\sqrt{3}}  \big[\bar c^d(x)i \gamma_5 u^d(x)\big]\big[\varepsilon^{abc} u^{a^T}(x)C\gamma_\alpha d^b(x)  \gamma^\alpha\gamma_5 c^c(x)\big]\nonumber\\
&
 \mp \sqrt{\frac{2}{3}} \big[\bar c^d(x)i \gamma_5 d^d(x)\big]  
 \big[\varepsilon^{abc} u^{a^T}(x) C\gamma_\alpha u^b(x) 
 \gamma^\alpha\gamma_5 c^c(x)\big] \, , \\
J_\mu^{\bar D \Sigma_c^*}(x)&=\frac{1}{\sqrt{3}}\mid \bar D^0(x) \Sigma_c^{+*}(x) \rangle \, \mp \sqrt{\frac{2}{3}}\mid \bar D^-(x) \Sigma_c^{*++}(x) \rangle  
= \frac{1}{\sqrt{3}}  \big[\bar c^d(x)i \gamma_5 u^d(x)\big]\big[\varepsilon^{abc} u^{a^T}(x)C\gamma_\mu d^b(x)  c^c(x)\big]\nonumber\\
&
 \mp \sqrt{\frac{2}{3}} \big[\bar c^d(x)i \gamma_5 d^d(x)\big]  
 \big[\varepsilon^{abc} u^{a^T}(x) C\gamma_\mu u^b(x) 
  c^c(x)\big] \, , \\  
J_\mu^{\bar D^* \Sigma_c}(x)& =\frac{1}{\sqrt{3}}\mid \bar D^{*0}(x) \Sigma_c^+(x) \rangle \, \mp \sqrt{\frac{2}{3}}\mid \bar D^{*-}(x) \Sigma_c^{++}(x) \rangle  
= \frac{1}{\sqrt{3}}  \big[\bar c^d(x)\gamma_\mu u^d(x)\big]\big[\varepsilon^{abc} u^{a^T}(x)C\gamma_\alpha d^b(x)  \gamma^\alpha\gamma_5 c^c(x)\big]\nonumber\\
&
 \mp \sqrt{\frac{2}{3}} \big[\bar c^d(x)\gamma_\mu d^d(x)\big]  
 \big[\varepsilon^{abc} u^{a^T}(x) C\gamma_\alpha u^b(x) 
 \gamma^\alpha\gamma_5 c^c(x)\big]\,, 
\end{align}
where  $a$, $b$, $c$ and $d$ are color indexes, and $C$ stands for the charge conjugation operator. As can be seen, $"\mp"$ appears in the explicit form of the interpolating currents written above. If the sign between these currents is minus, it couples to the isospin-1/2 states, while if it is plus, it couples to the isospin-3/2 states. As a result, in this study, the magnetic moments of both the isospin-1/2 states and the isospin-3/2 hidden-charm pentaquark states are determined.

In the hadronic representation, after inserting the full sets of hadronic states with the same quantum numbers as the interpolating currents and carrying out the Fourier integration over x, we acquire
 \begin{align}\label{edmn02}
\Pi^{Had}(p,q)&=\frac{\langle0\mid J^{\mathrm{P_c}}(x) \mid
{\mathrm{P_c}}(p, s) \rangle}{[p^{2}-m_{\mathrm{P_c}}^{2}]}
\langle {\mathrm{P_c}}(p, s)\mid
{\mathrm{P_c}}(p+q, s)\rangle_\gamma 
\frac{\langle {\mathrm{P_c}}(p+q, s)\mid
\bar J^{\mathrm{P_c}}(0) \mid 0\rangle}{[(p+q)^{2}-m_{\mathrm{P_c}}^{2}]}+ \cdots , \\
\Pi^{Had}_{\mu\nu}(p,q)&=\frac{\langle0\mid  J_{\mu}^{\mathrm{P_c^*}}(x)\mid
{\mathrm{P_c^*}}(p,s)\rangle}{[p^{2}-m_{{\mathrm{P_c^*}}}^{2}]}
\langle {\mathrm{P_c^*}}(p,s)\mid
{\mathrm{P_c^*}}(p+q,s)\rangle_\gamma 
\frac{\langle {\mathrm{P_c^*}}(p+q,s)\mid
\bar{J}_{\nu}^{\mathrm{P_c^*}}(0)\mid 0\rangle}{[(p+q)^{2}-m_{{\mathrm{P_c^*}}}^{2}]}+ \cdots .\label{Pc103}
\end{align}

For further calculations, the matrix elements in Eqs.~(\ref{edmn02}) and (\ref{Pc103}) are required and are written regarding the hadronic parameters as follows,
%
\begin{align}
\langle0\mid J^{\mathrm{P_c}}(x)\mid {\mathrm{P_c}}(p, s)\rangle=&\lambda_{\mathrm{P_c}} \gamma_5 \, u(p,s),\label{edmn04}\\
\langle {\mathrm{P_c}}(p+q, s)\mid\bar J^{\mathrm{P_c}}(0)\mid 0\rangle=&\lambda_{\mathrm{P_c}} \gamma_5 \, \bar u(p+q,s)\label{edmn004}
,\\
\langle {\mathrm{P_c}}(p, s)\mid {\mathrm{P_c}}(p+q, s)\rangle_\gamma &=\varepsilon^\mu\,\bar u(p, s)\bigg[\big[f_1(q^2)
+f_2(q^2)\big] \gamma_\mu +f_2(q^2)
\frac{(2p+q)_\mu}{2 m_{\mathrm{P_c}}}\bigg]\,u(p+q, s), \label{edmn005}\\
\langle0\mid J_{\mu}^{\mathrm{P_c^*}}(x)\mid {\mathrm{P_c^*}}(p,s)\rangle&=\lambda_{{\mathrm{P_c^*}}}u_{\mu}(p,s),\\
\langle {\mathrm{P_c^*}}(p+q,s)\mid
\bar{J}_{\nu}^{\mathrm{P_c^*}}(0)\mid 0\rangle &= \lambda_{{\mathrm{P_c^*}}}\bar u_{\nu}(p+q,s), \\
\langle {\mathrm{P_c^*}}(p,s)\mid {\mathrm{P_c^*}}(p+q,s)\rangle_\gamma &=-e\bar
u_{\mu}(p,s)\bigg[F_{1}(q^2)g_{\mu\nu}\eslash 
-
\frac{1}{2m_{{\mathrm{P_c^*}}}} 
\Big[F_{2}(q^2)g_{\mu\nu} 
+F_{4}(q^2)\frac{q_{\mu}q_{\nu}}{(2m_{{\mathrm{P_c^*}}})^2}\Big]\eslash\qslash
\nonumber\\
&+
F_{3}(q^2)\frac{1}{(2m_{{\mathrm{P_c^*}}})^2}q_{\mu}q_{\nu}\eslash \bigg] 
u_{\nu}(p+q,s),
\label{matelpar}
\end{align}
%
where $f_i$ and $F_i$'s are the Lorentz invariant form factors, the $u(p,s)$, $ u(p+q,s)$ and $\lambda_{{\mathrm{P_c}}}$ are the spinors and residue of the $\mathrm{P_c}$ state, respectively, and; the $u_{\mu}(p,s)$, $u_{\nu}(p+q,s)$ and $\lambda_{{\mathrm{P_c^*}}}$ are the spinors and residue of the $\mathrm{P_c^*}$ state, respectively. 

Combining the above equations, and summing over spins, we obtain the following expressions for the correlation functions of the hadronic representation:
\begin{align}
\label{edmn05}
\Pi^{Had}(p,q)=&\lambda^2_{\mathrm{P_c}}\gamma_5 \frac{\big(\pslash+m_{\mathrm{P_c}} \big)}{[p^{2}-m_{{\mathrm{P_c}}}^{2}]}\varepsilon^\mu \bigg[\big[f_1(q^2) %
+f_2(q^2)\big] \gamma_\mu
+f_2(q^2)\, \frac{(2p+q)_\mu}{2 m_{\mathrm{P_c}}}\bigg]  \gamma_5 
\frac{\big(\pslash+\qslash+m_{\mathrm{P_c}}\big)}{[(p+q)^{2}-m_{{\mathrm{P_c}}}^{2}]}, \\
\Pi^{Had}_{\mu\nu}(p,q)&=\frac{\lambda_{_{{\mathrm{P_c^*}}}}^{2}}{[(p+q)^{2}-m_{_{{\mathrm{P_c^*}}}}^{2}][p^{2}-m_{_{{\mathrm{P_c^*}}}}^{2}]} 
\bigg[  g_{\mu\nu}\pslash\eslash\qslash \,F_{1}(q^2) 
-m_{{\mathrm{P_c^*}}}g_{\mu\nu}\eslash\qslash\,F_{2}(q^2)
-
\frac{F_{3}(q^2)}{4m_{{\mathrm{P_c^*}}}}q_{\mu}q_{\nu}\eslash\qslash \nonumber\\
&
-
\frac{F_{4}(q^2)}{4m_{{\mathrm{P_c^*}}}^3}(\varepsilon.p)q_{\mu}q_{\nu}\pslash\qslash 
+
\cdots 
\bigg]. \label{final phenpart}
\end{align}

To calculate the magnetic form factor, $F_M(Q^2)(G_M(Q^2)$, of these pentaquark states, these form factors must be written in terms of the form factors $f_i(Q^2)$ and$F_i(Q^2)$, and these expressions are given as follows
\begin{align}
\label{edmn07}
F_M(q^2) &= f_1(q^2) + f_2(q^2),\\
G_{M}(q^2) &= \left[ F_1(q^2) + F_2(q^2)\right] ( 1+ \frac{4}{5}
\lambda ) -\frac{2}{5} \left[ F_3(q^2)  \right]
+\left[
F_4(q^2)\right] \lambda \left( 1 + \lambda \right),
\end{align}  where $F_M(q^2)$ and $G_M(q^2)$ are magnetic form factor for spin-1/2 and spin-3/2 states, respectively and;  $\lambda
= -\frac{q^2}{4m^2_{{\mathrm{P_c^*}}}}$.  Using the above expressions, we can obtain the electromagnetic form factors of these pentaquarks. However, since we are dealing with a real photon ($q^2 =0$), we can obtain these form factors in terms of the magnetic moment, and their explicit expressions are as follows
\begin{align}
\label{edmn08}
\mu_{\mathrm{P_c}} &= \frac{ e}{2\, m_{\mathrm{P_c}}} \,F_M(0),\\~~~~
\mu_{{\mathrm{P_c^*}}}&=\frac{e}{2m_{{\mathrm{P_c^*}}}}G_{M}(0),
\end{align}
where $F_M(0)= f_1(0)+f_2(0)$ and $G_{M}(0)=F_1(0)+F_2(0)$.  While obtaining the aforementioned expressions, the Lorentz structures $\eslash \qslash$, $g_{\mu\nu} \pslash \eslash \qslash$ and $g_{\mu\nu}  \eslash \qslash$  have been selected for the $(f_1(0)+f_2(0))$, $F_1(0)$ and $F_2(0)$ form factors, respectively. The rationale behind prioritizing these Lorentz structures is that they encompass more powers of momentum, which serves to optimize the convergence of the operator product expansion, and hence causes a more reliable derivation of the magnetic moments of hidden-charm pentaquarks.

On the QCD side, the correlation function is achieved in the deep Euclidean region by operator product expansion (OPE). To proceed, we should derive the correlation function employing the light and heavy quark propagators and the distribution amplitudes of the photon.  Consequently, the results of the contractions  for the $\bar D \Sigma_c$, $\bar D^{*} \Sigma_c$, and $\bar D \Sigma_c^{*}$ pentaquarks are listed as follows
\begin{align}
\label{QCD1}
\Pi^{\rm{QCD}-\mathrm{\bar D \Sigma_c}}(p,q)&= \frac{i}{3}\varepsilon^{abc} \varepsilon^{a^{\prime}b^{\prime}c^{\prime}}\, \int d^4x \, e^{ip\cdot x} 
\nonumber\\
& 
 \langle 0\mid \Big\{ \mbox{Tr}\Big[\gamma_5 S_{u}^{dd^\prime}(x) \gamma_5  S_{c}^{d^\prime d}(-x)\Big]  
\mbox{Tr}\Big[\gamma_{\alpha} S_d^{bb^\prime}(x) \gamma_{\beta}  
  \widetilde S_{u}^{aa^\prime}(x)\Big]
(\gamma^{\alpha}\gamma_5 S_{c}^{cc^\prime}(x) \gamma_5  \gamma^{\beta})
 \nonumber\\
&     
\mp  \mbox{Tr}\Big[\gamma_5 S_{u}^{da^\prime}(x) \gamma_{\beta} \widetilde S_{d}^{bb^\prime}(x)
   \gamma_{\alpha}  S_u^{ad^\prime} \gamma_5(x) S_{c}^{d^\prime d}(-x)\Big] 
(\gamma^{\alpha}\gamma_5 S_{c}^{cc^\prime}(x) \gamma_5  \gamma^{\beta})\nonumber\\
&
  +2 \mbox{Tr}\Big[\gamma_5 S_{d}^{dd^\prime}(x) \gamma_5  S_{c}^{d^\prime d}(-x)\Big]  
\mbox{Tr}\Big[\gamma_{\alpha} S_u^{bb^\prime}(x) 
 \gamma_{\beta} \widetilde S_{u}^{aa^\prime}(x)\Big]
(\gamma^{\alpha}\gamma_5 S_{c}^{cc^\prime}(x) \gamma_5  \gamma^{\beta})
 \nonumber\\
&  \mp 2  \mbox{Tr}\Big[\gamma_5 S_{d}^{dd^\prime}(x) \gamma_5  S_{c}^{d^\prime d}(-x)\Big]   
 \mbox{Tr}\Big[\gamma^{\alpha} S_{u}^{ba^\prime}(x) 
 \gamma_{\beta} \widetilde S_u^{ab^\prime}(x) \Big] 
(\gamma_{\alpha}\gamma_5 S_{c}^{cc^\prime}(x) \gamma_5  \gamma^{\beta})
 \Big\}
\mid 0 \rangle _\gamma \,,
\end{align}
\begin{align}
\Pi_{\mu\nu}^{\rm{QCD}-\mathrm{\bar D \Sigma_c^*}}(p,q)&= \frac{i}{3}\varepsilon^{abc} \varepsilon^{a^{\prime}b^{\prime}c^{\prime}}\, \int d^4x \, e^{ip\cdot x} 
\nonumber\\
& 
 \langle 0\mid \Big\{ \mbox{Tr}\Big[\gamma_5 S_{u}^{dd^\prime}(x) \gamma_5  S_{c}^{d^\prime d}(-x)\Big]  
\mbox{Tr}\Big[\gamma_{\mu} S_d^{bb^\prime}(x) \gamma_{\nu}  
  \widetilde S_{u}^{aa^\prime}(x)\Big]
 S_{c}^{cc^\prime}(x) 
 \nonumber\\
&     
\mp  \mbox{Tr}\Big[\gamma_5 S_{u}^{da^\prime}(x) \gamma_{\nu}   \widetilde S_{d}^{bb^\prime}(x)
   \gamma_{\mu}  S_u^{ad^\prime} \gamma_5(x) S_{c}^{d^\prime d}(-x)\Big] 
 S_{c}^{cc^\prime}(x) \nonumber\\
&
  +2 \mbox{Tr}\Big[\gamma_5 S_{d}^{dd^\prime}(x) \gamma_5  S_{c}^{d^\prime d}(-x)\Big]  
\mbox{Tr}\Big[\gamma_{\mu} S_u^{bb^\prime}(x) 
 \gamma_{\nu} \widetilde S_{u}^{aa^\prime}(x)\Big]
 S_{c}^{cc^\prime}(x) 
 \nonumber\\
&  \mp 2  \mbox{Tr}\Big[\gamma_5 S_{d}^{dd^\prime}(x) \gamma_5  S_{c}^{d^\prime d}(-x)\Big]   
 \mbox{Tr}\Big[\gamma_{\mu} S_{u}^{ba^\prime}(x) 
 \gamma_{\nu} \widetilde S_u^{ab^\prime}(x) \Big] 
 S_{c}^{cc^\prime}(x) 
 \Big\}
\mid 0 \rangle _\gamma \,,\label{QCD2}
\end{align}
\begin{align}
\Pi_{\mu\nu}^{\rm{QCD}-\mathrm{\bar D^* \Sigma_c}}(p,q)&= \frac{i}{3}\varepsilon^{abc} \varepsilon^{a^{\prime}b^{\prime}c^{\prime}}\, \int d^4x \, e^{ip\cdot x} 
\nonumber\\
& 
 \langle 0\mid \Big\{ \mbox{Tr}\Big[\gamma_{\alpha} S_{u}^{dd^\prime}(x) \gamma_{\beta}  S_{c}^{d^\prime d}(-x)\Big]  
\mbox{Tr}\Big[\gamma_{\mu} S_d^{bb^\prime}(x) \gamma_{\nu}  
  \widetilde S_{u}^{aa^\prime}(x)\Big]
(\gamma^{\alpha}\gamma_5 S_{c}^{cc^\prime}(x) \gamma_5  \gamma^{\beta})
 \nonumber\\
&     
\mp  \mbox{Tr}\Big[\gamma_{\alpha} S_{u}^{da^\prime}(x) \gamma_{\nu}   \widetilde S_{d}^{bb^\prime}(x)
   \gamma_{\mu}  S_u^{ad^\prime} \gamma_{\beta}(x) S_{c}^{d^\prime d}(-x)\Big] 
 (\gamma^{\alpha}\gamma_5 S_{c}^{cc^\prime}(x) \gamma_5  \gamma^{\beta}) \nonumber\\
&
  +2 \mbox{Tr}\Big[\gamma_{\alpha} S_{d}^{dd^\prime}(x) \gamma_{\beta}  S_{c}^{d^\prime d}(-x)\Big]  
\mbox{Tr}\Big[\gamma_{\mu} S_u^{bb^\prime}(x) 
 \gamma_{\nu} \widetilde S_{u}^{aa^\prime}(x)\Big]
 (\gamma^{\alpha}\gamma_5 S_{c}^{cc^\prime}(x) \gamma_5  \gamma^{\beta}) 
 \nonumber\\
&  \mp 2  \mbox{Tr}\Big[\gamma_{\alpha} S_{d}^{dd^\prime}(x) \gamma_{\beta}  S_{c}^{d^\prime d}(-x)\Big]   
 \mbox{Tr}\Big[\gamma_{\mu} S_{u}^{ba^\prime}(x) 
 \gamma_{\nu} \widetilde S_u^{ab^\prime}(x) \Big] 
 (\gamma^{\alpha}\gamma_5 S_{c}^{cc^\prime}(x) \gamma_5  \gamma^{\beta}) 
 \Big\}
\mid 0 \rangle _\gamma \,, \label{QCD3}
\end{align}
where   
$\widetilde{S}_{c(q)}^{ij}(x)=CS_{c(q)}^{ij\mathrm{T}}(x)C$. The heavy and the light quark propagators are designated as $S_{c}(x)$ and $S_{q}(x)$, respectively, and are defined as follows~\cite{Yang:1993bp, Belyaev:1985wza}:
\begin{align}
\label{edmn13}
S_{q}(x)&= \frac{1}{2 \pi x^2}\Big(i \frac{\xslash}{x^2}- \frac{m_q}{2}\Big)
- \frac{\langle \bar qq \rangle }{12} \Big(1-i\frac{m_{q} \xslash}{4}   \Big)
- \frac{ \langle \bar qq \rangle }{192}
m_0^2 x^2  \Big(1
  -i\frac{m_{q} \xslash}{6}   \Big)
-\frac {i g_s }{32 \pi^2 x^2} ~G^{\mu \nu} (x) 
\Big[\rlap/{x} 
\sigma_{\mu \nu} +  \sigma_{\mu \nu} \rlap/{x}
 \Big],\\
\nonumber\\
S_{c}(x)&=\frac{m_{c}^{2}}{4 \pi^{2}} \Bigg[ \frac{K_{1}\Big(m_{c}\sqrt{-x^{2}}\Big) }{\sqrt{-x^{2}}}
+i\frac{{\xslash}~K_{2}\Big( m_{c}\sqrt{-x^{2}}\Big)}
{(\sqrt{-x^{2}})^{2}}\Bigg]
-\frac{g_{s}m_{c}G^{\mu \nu }(x)}{32 \pi ^{2}} \Bigg[ (\sigma _{\mu \nu }{\xslash}
  +{\xslash}\sigma _{\mu \nu }) 
    \frac{K_{1}\Big( m_{c}\sqrt{-x^{2}}\Big) }{\sqrt{-x^{2}}}
   \nonumber\\
  &
 +2\sigma_{\mu \nu }K_{0}\Big( m_{c}\sqrt{-x^{2}}\Big)\Bigg], 
 \label{edmn14}
\end{align}%
where $\langle \bar qq \rangle$ represents the light-quark condensate, $m_0$ is defined through the quark-gluon mixed condensate $ m_0^2= \langle 0 \mid \bar  q\, g_s\, \sigma_{\mu\nu}\, G^{\mu\nu}\, q \mid 0 \rangle / \langle \bar qq \rangle $, $G^{\mu\nu}$ corresponds to the gluon field strength tensor,  and $K_i$'s refer to the modified Bessel functions of the second kind, respectively.    The first terms of these propagators correspond to the perturbative or free part and the remaining terms belong to the non-perturbative or interacting parts.

The correlation functions in the QCD representation contain two different contributions:
\begin{itemize}
 \item perturbative contributions, where the photon is radiated at short-distances, 
 \item non-perturbative contributions, where the photon is radiated at long-distances.
\end{itemize}

These contributions should be calculated for the analysis to be complete and reliable. To calculate the perturbative contributions, one of the quark propagators perturbatively interacts with the photon and is substituted by 

\begin{align}
\label{free}
S^{free}(x) \rightarrow \int d^4y\, S^{free} (x-z)\,\rlap/{\!A}(z)\, S^{free} (z)\,,
\end{align}
and the remaining propagators in Eqs.~(\ref{QCD1})-(\ref{QCD3}) have been taken into account to be free.  The non-perturbative effects are derived by substituting one of the light quark propagators in Eqs.~(\ref{QCD1})-(\ref{QCD3}) as follows

 \begin{align}
\label{edmn21}
S_{\alpha\beta}^{ab}(x) \rightarrow -\frac{1}{4} \Big[\bar{q}^a(x) \Gamma_i q^b(0)\Big]\Big(\Gamma_i\Big)_{\alpha\beta},
\end{align}
and in Eqs.~(\ref{QCD1})-(\ref{QCD3}) the surviving propagators have been taken into account as full propagators. Here $\Gamma_i = \{\textbf{1}, \gamma_5, \gamma_\mu, i\gamma_5 \gamma_\mu, \sigma_{\mu\nu}/2\}$. When Eq.~(\ref{edmn21}) is plugged into Eqs.~(\ref{QCD1})-(\ref{QCD3}), matrix elements of the $\langle \gamma(q)\vel \bar{q}(x) \Gamma_i G_{\alpha\beta}q(0) \ver 0\rangle$ and $\langle \gamma(q)\vel \bar{q}(x) \Gamma_i q(0) \ver 0\rangle$  appear which are described regarding the distribution amplitudes of the photon and characterize the interaction of photons with quark fields at large distance.  These matrix elements, characterized by photon wave functions with specific twists, are very important for the computation of non-perturbative effects (for details on the distribution amplitudes of the photon, see Ref.~\cite{Ball:2002ps}).    
After the technical and tedious procedures described above, the QCD representation of the magnetic moment analysis is obtained.

The QCD light-cone sum rules for the magnetic moments of the $\bar D \Sigma_c$, $\bar D^{*} \Sigma_c$, and $\bar D \Sigma_c^{*}$ pentaquarks are achieved by relating correlation function expressions employing QCD quantities to expressions using hadron quantities by means of quark-hadron duality. To dominate the contributions of the continuum and the higher states and to boost the contribution of the ground state, we perform the continuum subtraction and the Borel transformation procedures based on the standard methodology of the QCD light-cone sum rules method. The results obtained using all of the above procedures for magnetic moments are as follows:

\begin{align}
\label{edmn15}
&\mu_{\bar D \Sigma_c} \,\lambda^2_{\bar D \Sigma_c}\, m_{\bar D \Sigma_c}  =e^{\frac{m^2_{\bar D \Sigma_c}}{\rm{M^2}}}\, \Delta_1^{\rm{QCD}} (\rm{M^2},\rm{s_0}),\\
\nonumber\\
&\mu_{\bar D \Sigma_c^*} \,\lambda^2_{\bar D \Sigma_c^*}\, m_{\bar D \Sigma_c^*}  =e^{\frac{m^2_{\bar D \Sigma_c^*}}{\rm{M^2}}}\, \Delta_2^{\rm{QCD}} (\rm{M^2},\rm{s_0}),\\
\nonumber\\
&\mu_{\bar D^* \Sigma_c} \,\lambda^2_{\bar D^* \Sigma_c}\, m_{\bar D^* \Sigma_c} =e^{\frac{m^2_{\bar D^* \Sigma_c}}{\rm{M^2}}}\, \Delta_3^{\rm{QCD}} (\rm{M^2},\rm{s_0}).
\end{align}
For simplicity, we list only the explicit form of the $\Delta_1^{\rm{QCD}}(\rm{M^2},\rm{s_0})$ function in the Appendix.

It should be noted that the Borel transformations in the aforementioned equations are carried out following the following equations.

\begin{align}
 \mathcal{B}\bigg\{ \frac{1}{\big[ [p^2-m^2_i][(p+q)^2-m_f^2] \big]}\bigg\} \rightarrow e^{-m_i^2/M_1^2-m_f^2/M_2^2}
\end{align}
in the hadronic side, and 

\begin{align}
 \mathcal{B}\bigg\{ \frac{1}{\big(m^2- \bar u p^2-u(p+q)^2\big)^{\alpha}}\bigg\} \rightarrow (M^2)^{(2-\alpha)} \delta (u-u_0)e^{-m^2/M^2},
\end{align}
 in the QCD side,  where we use
 
\begin{align*}
 {M^2}= \frac{M_1^2 M_2^2}{M_1^2+M_2^2}, \\
 u_0= \frac{M_1^2}{M_1^2+M_2^2}.
\end{align*}
Here $ M_1^2 $ and $ M_2^2 $ are the Borel parameters in the initial and final states, respectively. Since we have the same hidden-charm pentaquarks in the initial and final states, therefore we can set, M$_1^2$ = M$_2^2$= 2M$^2$ and $u_0 = 1/2 $, which leads to the single dispersion approximation being sufficient to suppress higher states and continuum contributions. Further details regarding this procedure can be found in Ref.~\cite{Ozdem:2024dbq}.

\end{widetext}

\section{Results and discussions}\label{numerical}

We proceed to the numerical analysis of the \rm{QCD} sum rules for the magnetic moments after their extraction in the previous section. They include several input parameters such as the masses of the quarks used in the calculations, the quark, gluon, and quark-gluon mixed condensates, the masses and residues of the hadrons, etc., and numerical values of these parameters are given in Table \ref{inputparameter}. The photon distribution amplitudes and their input parameters, which are required for further analysis, are borrowed from Ref.~\cite{Ball:2002ps}.

\begin{widetext}

 \begin{table}[t]
	\addtolength{\tabcolsep}{10pt}
	\caption{Parameters employed as inputs in the calculations
	.}
	\label{inputparameter}
\begin{tabular}{l|c|c|cccc}
               \hline\hline
Parameters & Values&Unit&References \\
                                        \hline\hline
$m_c$&$ 1.27 \pm 0.02$&GeV  &\cite{Workman:2022ynf}
                        \\
$m^{I=1/2}_{\bar D \Sigma_c}$&$  4.31^{+0.07}_{-0.07}$&GeV&  \cite{Wang:2022ltr}
                        \\
$m^{I=3/2}_{\bar D \Sigma_c}  $&$ 4.33^{+0.09}_{-0.08}$&GeV& \cite{Wang:2022ltr}
                        \\
$m^{I=1/2}_{\bar D \Sigma_c^*}$&$ 4.38^{+0.07}_{-0.07}$&GeV& \cite{Wang:2022ltr}
                        \\
$m^{I=3/2}_{\bar D \Sigma_c^*}  $&$ 4.41^{+0.08}_{-0.08}$&GeV&  \cite{Wang:2022ltr}
                       \\
$m^{I=1/2}_{\bar D^* \Sigma_c} $&$  4.44^{+0.07}_{-0.08}$&GeV& \cite{Wang:2022ltr}
                       \\
$m^{I=3/2}_{\bar D^* \Sigma_c} $&$ 4.47^{+0.0}_{-0.09}$&GeV& \cite{Wang:2022ltr}
                       \\
$m_0^{2} $&$ 0.8 \pm 0.1 $&\,\,GeV$^2$  &\cite{Ioffe:2005ym}
                       \\
$f_{3\gamma} $&$ -0.0039 $&\,\,GeV$^2$ &\cite{Ball:2002ps}
                       \\
$\chi $&$ -2.85 \pm 0.5 $&\,\,\,\,\,GeV$^{-2}$ &\cite{Rohrwild:2007yt}
                       \\
$\langle \bar qq\rangle $&$ (-0.24 \pm 0.01)^3 $&\,\,GeV$^3$& \cite{Ioffe:2005ym}
                       \\
$ \langle g_s^2G^2\rangle  $&$ 0.48 \pm 0.14 $&\,\,GeV$^4$ &\cite{Narison:2018nbv}
                       \\
$\lambda^{I=1/2}_{\bar D \Sigma_c}  $&$ (3.25^{+0.43}_{-0.41})\times 10^{-3} $&\,\,GeV$^6$& \cite{Wang:2022ltr}
                       \\
$ \lambda^{I=3/2}_{\bar D \Sigma_c}   $&$(1.97^{+0.28}_{-0.26})\times 10^{-3} $&\,\,GeV$^6$& \cite{Wang:2022ltr}
                       \\
$ \lambda^{I=1/2}_{\bar D \Sigma_c^*}   $&$ (1.97^{+0.46}_{-0.24})\times 10^{-3} $&\,\,GeV$^6$ &\cite{Wang:2022ltr}
                       \\
$ \lambda^{I=3/2}_{\bar D \Sigma_c^*}  $&$ (1.24^{+0.17}_{-0.16})\times 10^{-3} $&\,\,GeV$^6$& \cite{Wang:2022ltr}
                       \\
$\lambda^{I=1/2}_{\bar D^* \Sigma_c}   $&$(3.60^{+0.47}_{-0.44})\times 10^{-3} $&\,\,GeV$^6$& \cite{Wang:2022ltr}
                       \\
$\lambda^{I=3/2}_{\bar D^* \Sigma_c}   $&$(2.31^{+0.33}_{-0.31})\times 10^{-3} $&\,\,GeV$^6$& \cite{Wang:2022ltr}
                       \\
                                      \hline\hline
 \end{tabular}
\end{table}

\end{widetext}

There are two additional parameters, the Borel parameter $\rm{M^2}$ and the threshold parameter $\rm{s_0}$, in addition to the parameters listed above. They are obtained from the analysis of the results based on the standard constraints of the QCD sum rule method. 
The upper and lower limits of the $\rm{M^2}$ are determined from the convergence of the OPE (CVG) and the pole dominance (PC). To characterize these constraints, it is useful to use the following expressions:
\begin{align}
 \mbox{PC} &=\frac{\Delta_i (\rm{M^2},\rm{s_0})}{\Delta_i (\rm{M^2},\infty)} \geq 30\%,\\
 \nonumber\\
 \mbox{CVG} &=\frac{\Delta_i^{\mbox{DimN}} (\rm{M^2},\rm{s_0})}{\Delta_i (\rm{M^2},\rm{s_0})} \leq 5\%,
 \end{align}
 where $\Delta_i^{\mbox{DimN}} (\rm{M^2},\rm{s_0})$ represent the highest dimensional terms in the  OPE of the $\Delta_i (\rm{M^2},\rm{s_0})$. In the OPE side of our analysis, the highest dimensional terms are dimension 10 ($\langle g_s^2G^2\rangle  \langle \bar q q\rangle^2$) and 11 ($f_{3\gamma} m_0^2$ $\langle g_s^2G^2\rangle  \langle \bar q q\rangle$). Consequently, the CVG analysis has been performed by considering the DimN expression in the form of $\rm{Dim(10+11)}$, and the CVG results in the table \ref{parameter} reflect this.  The QCD side of the analysis also encompasses the possible combinations of condensate expressions, including $\langle g_s^2G^2\rangle  \langle \bar q q\rangle^2$, $m_0^2$ $\langle g_s^2G^2\rangle  \langle \bar q q\rangle$, $\langle g_s^2G^2\rangle  \langle \bar q q\rangle$, $ \langle \bar q q\rangle^2$,  $\langle g_s^2G^2\rangle$, and $\langle \bar q q\rangle$. 

\begin{widetext}

\begin{figure}[htb!]
\subfloat[]{\includegraphics[width=0.4\textwidth]{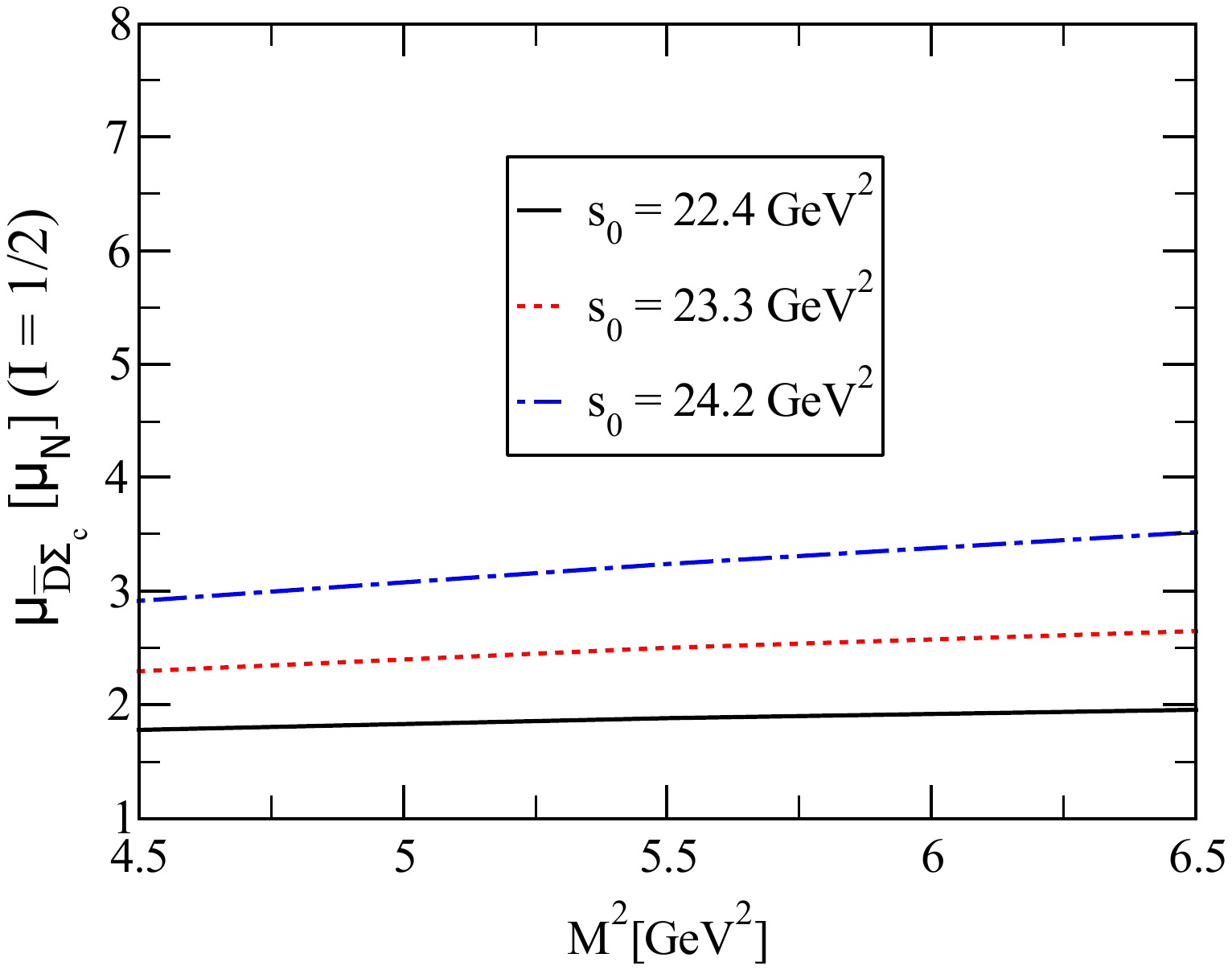}}~~~~~~~~~~~~~~~~~~
\subfloat[]{\includegraphics[width=0.4\textwidth]{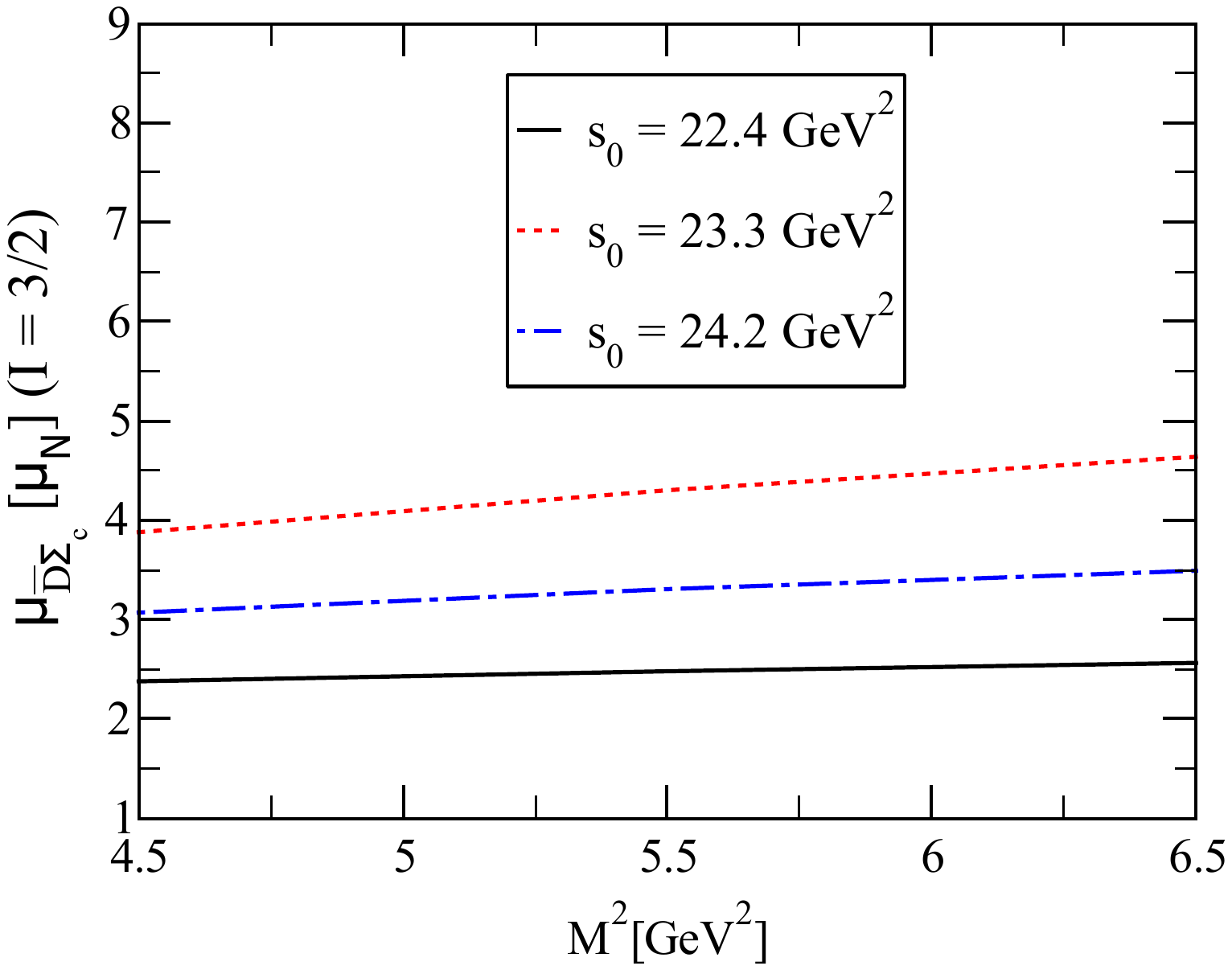}}\\
\subfloat[]{\includegraphics[width=0.4\textwidth]{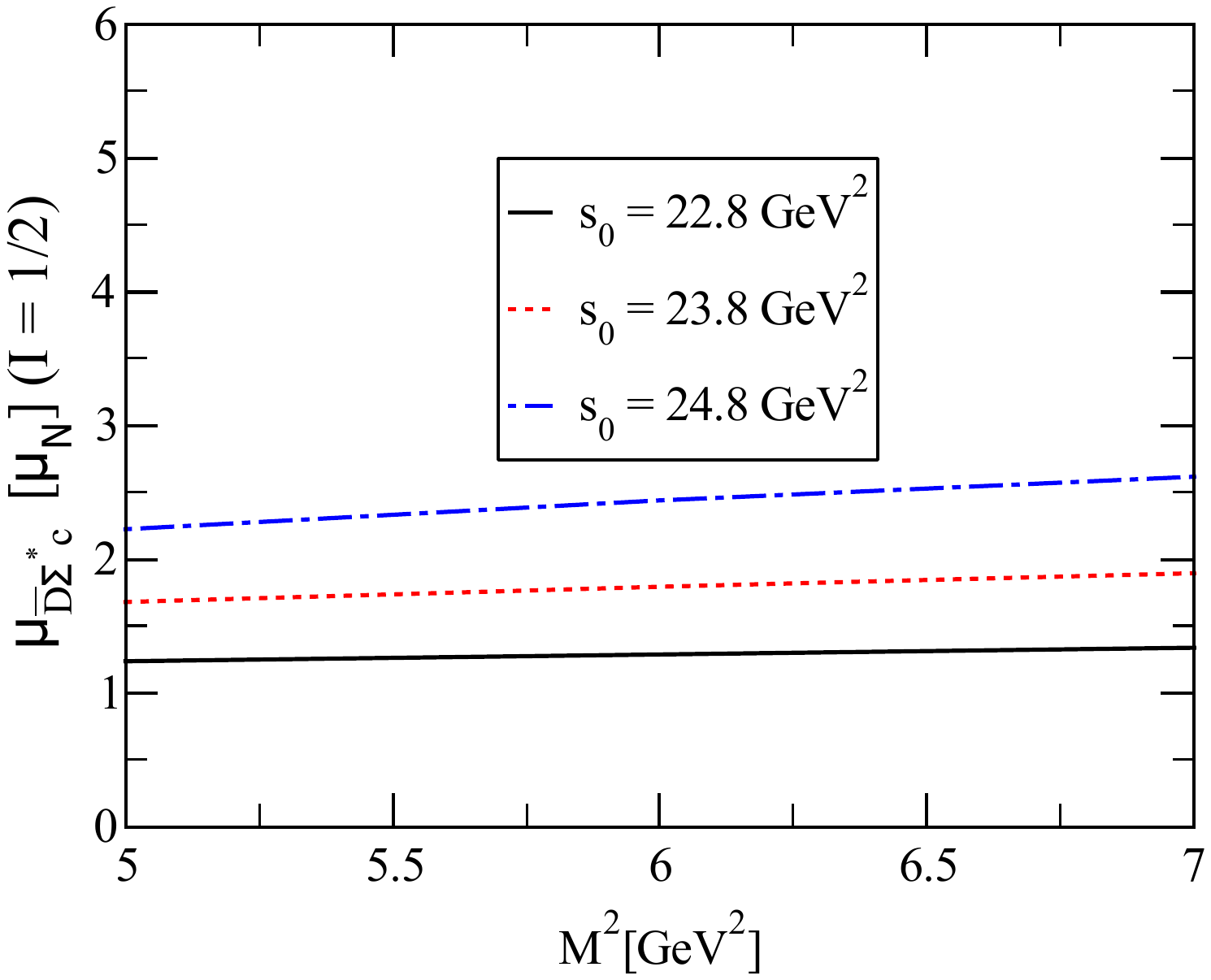}}~~~~~~~~~~~~~~~~~~
\subfloat[]{\includegraphics[width=0.4\textwidth]{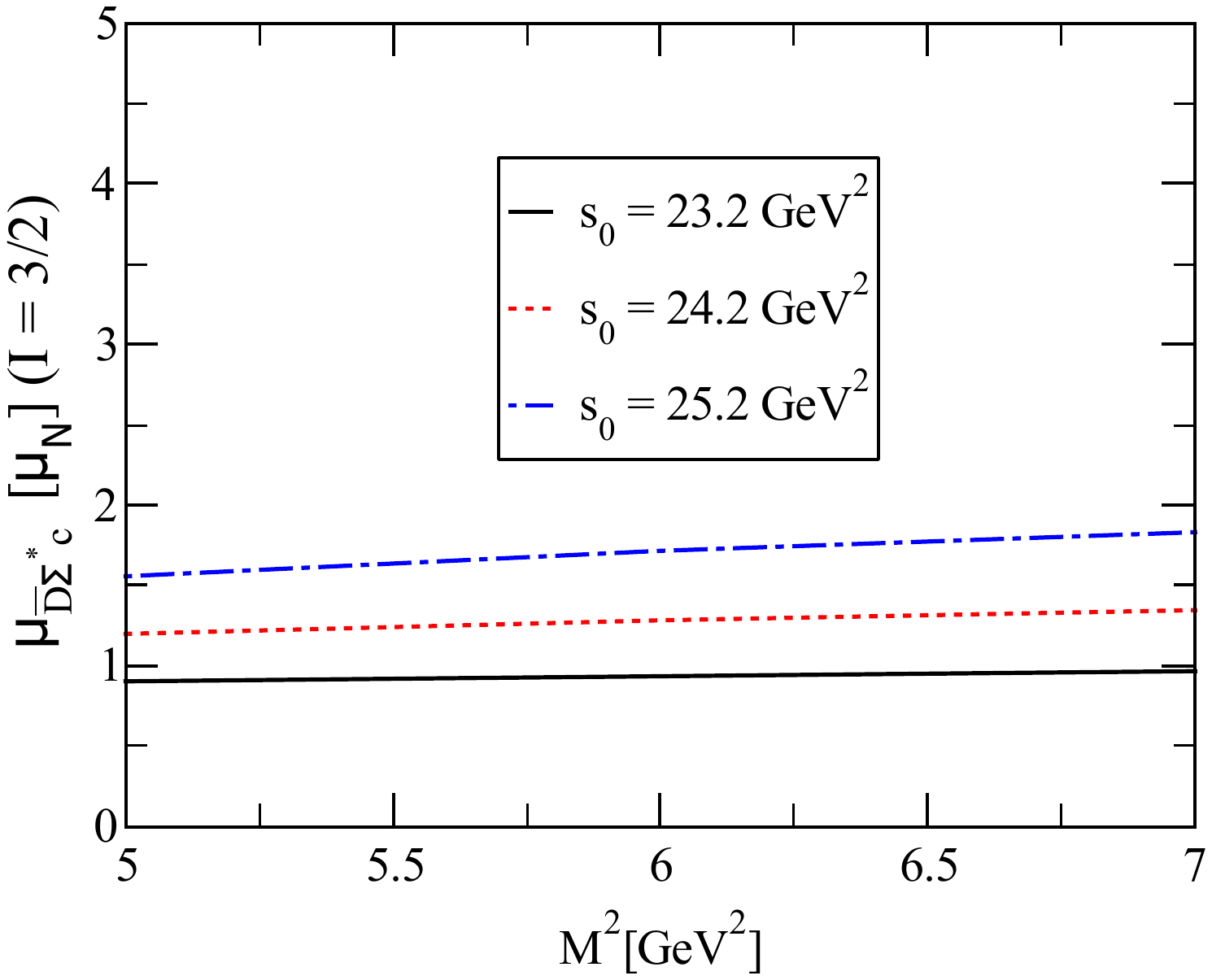}}\\
\subfloat[]{\includegraphics[width=0.4\textwidth]{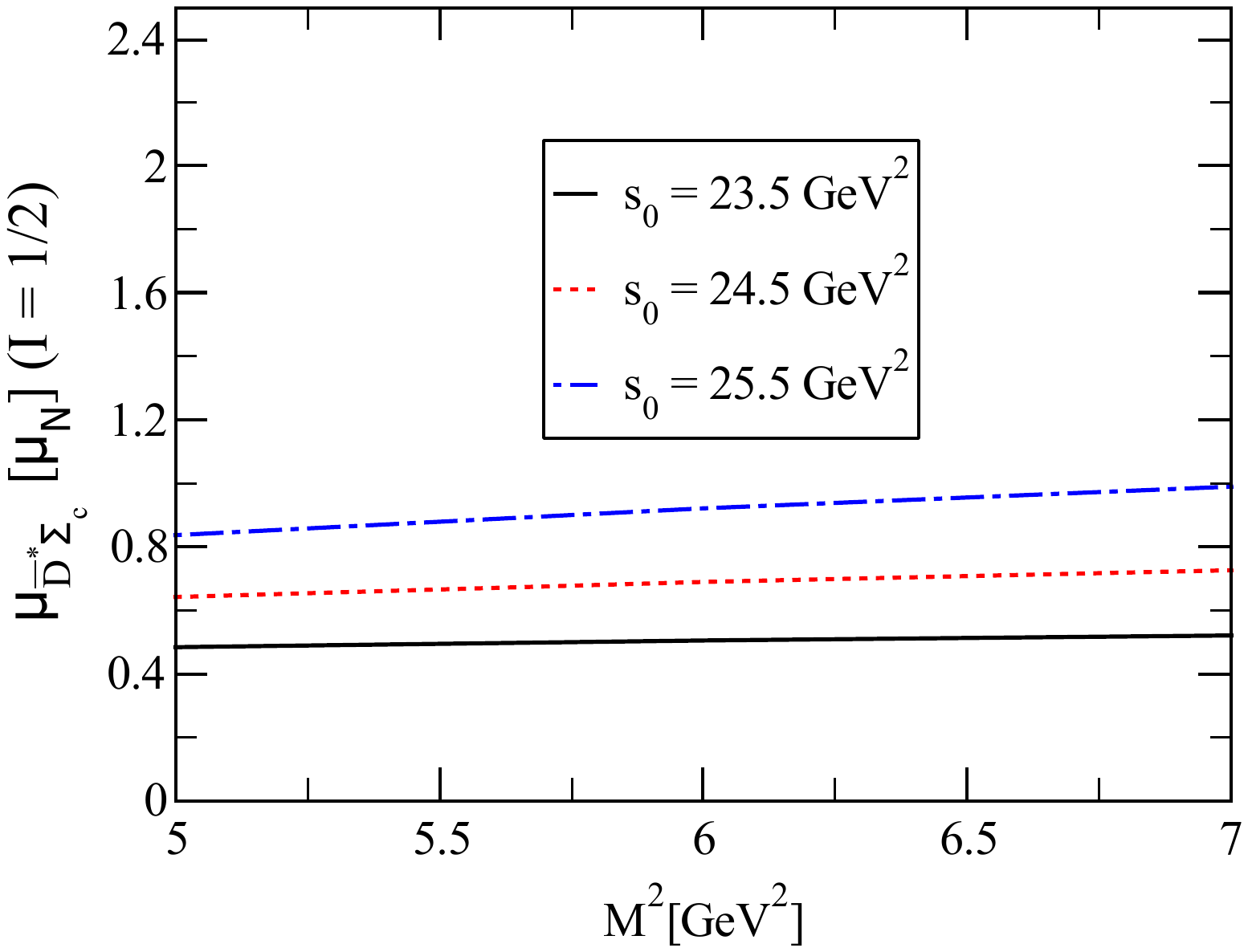}}~~~~~~~~~~~~~~~~~~
\subfloat[]{\includegraphics[width=0.4\textwidth]{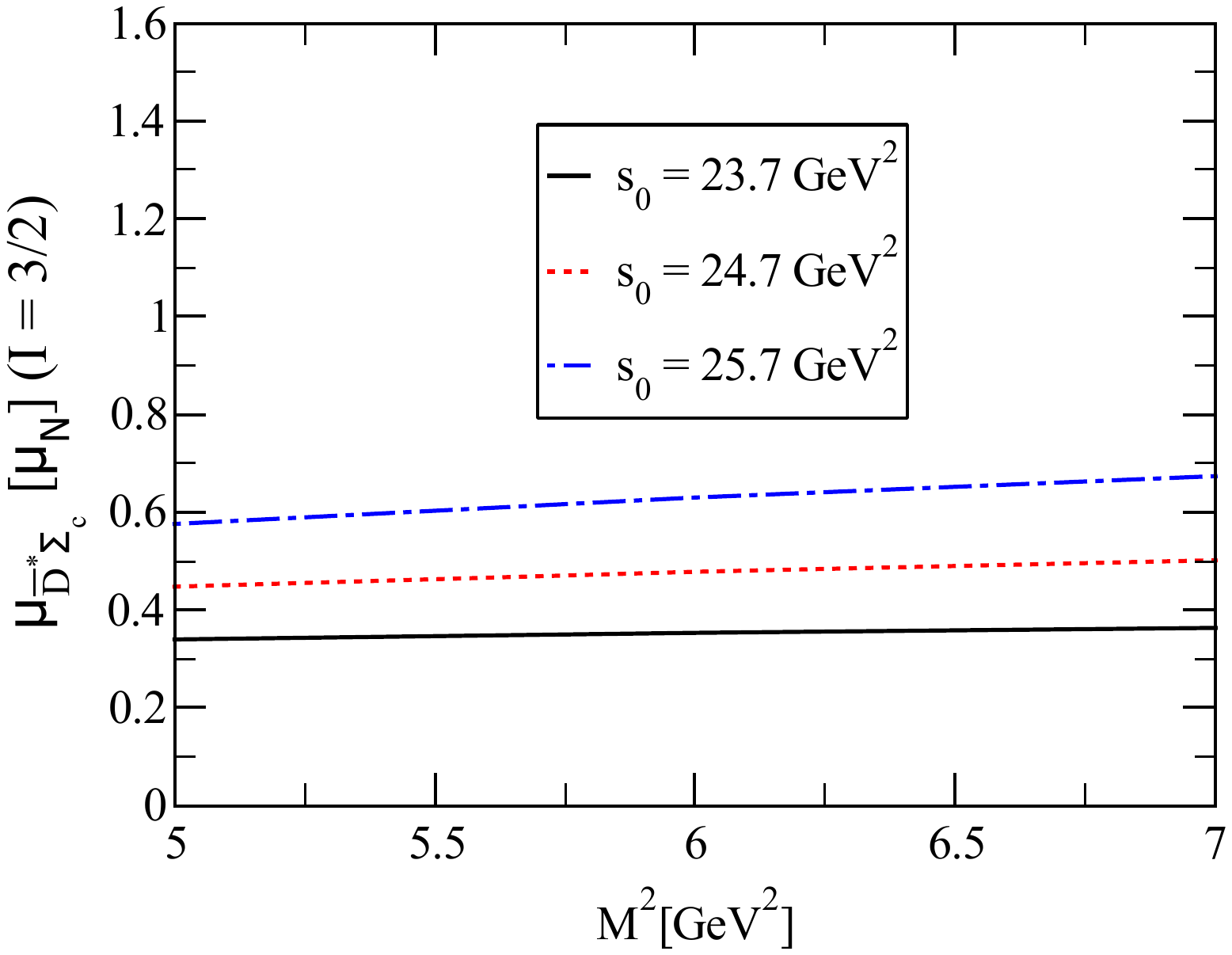}}
 \caption{The magnetic  moments of the isospin eigenstate $\bar D \Sigma_c$, $\bar D^{*} \Sigma_c$, and $\bar D \Sigma_c^{*}$ pentaquarks versus $\rm{M^2}$ at three different values of $\rm{s_0}$.}
 \label{Msqfig1}
  \end{figure}

  \end{widetext}

\begin{widetext}

\begin{figure}[htb!]
\subfloat[]{\includegraphics[width=0.4\textwidth]{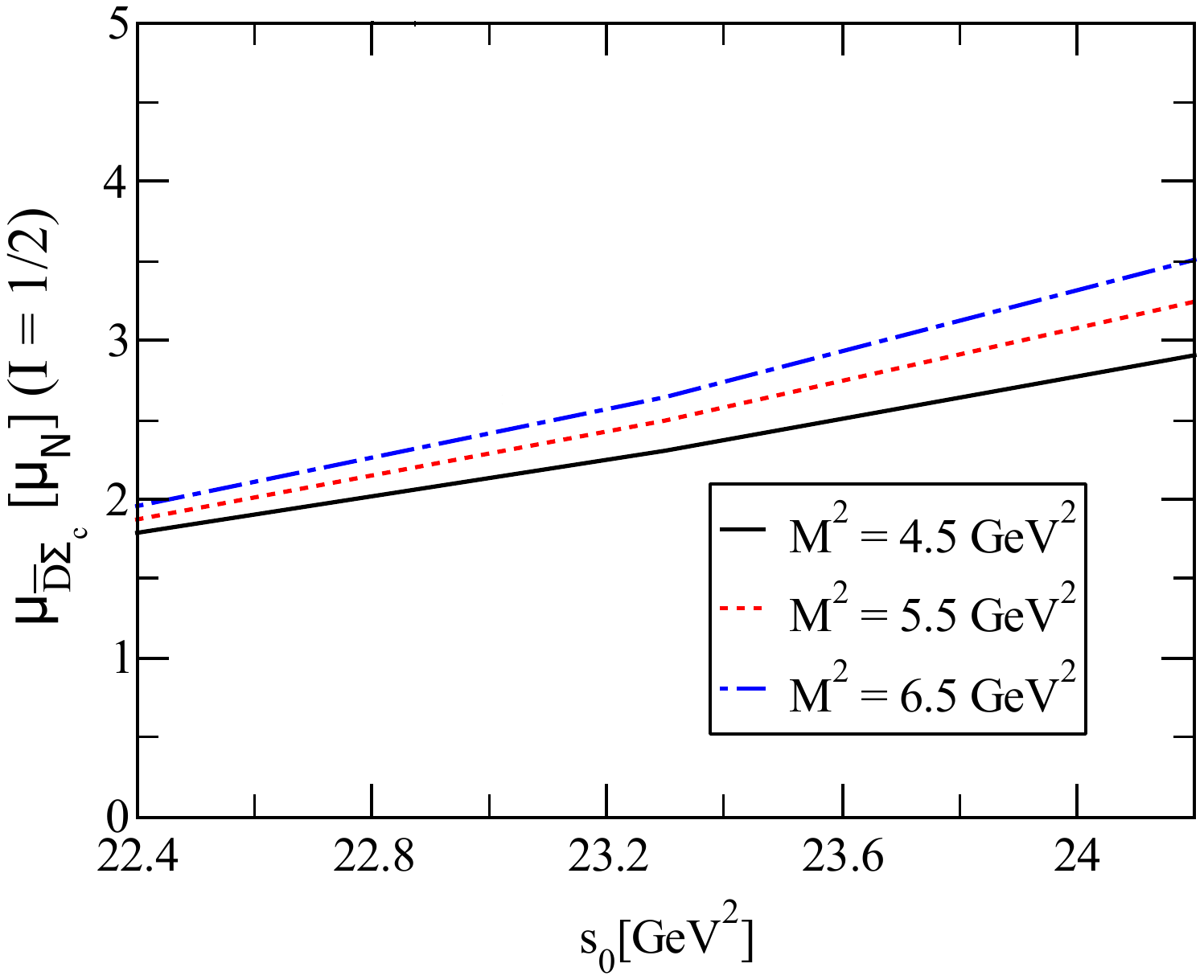}}~~~~~~~~~~~~~~~~~~
\subfloat[]{\includegraphics[width=0.4\textwidth]{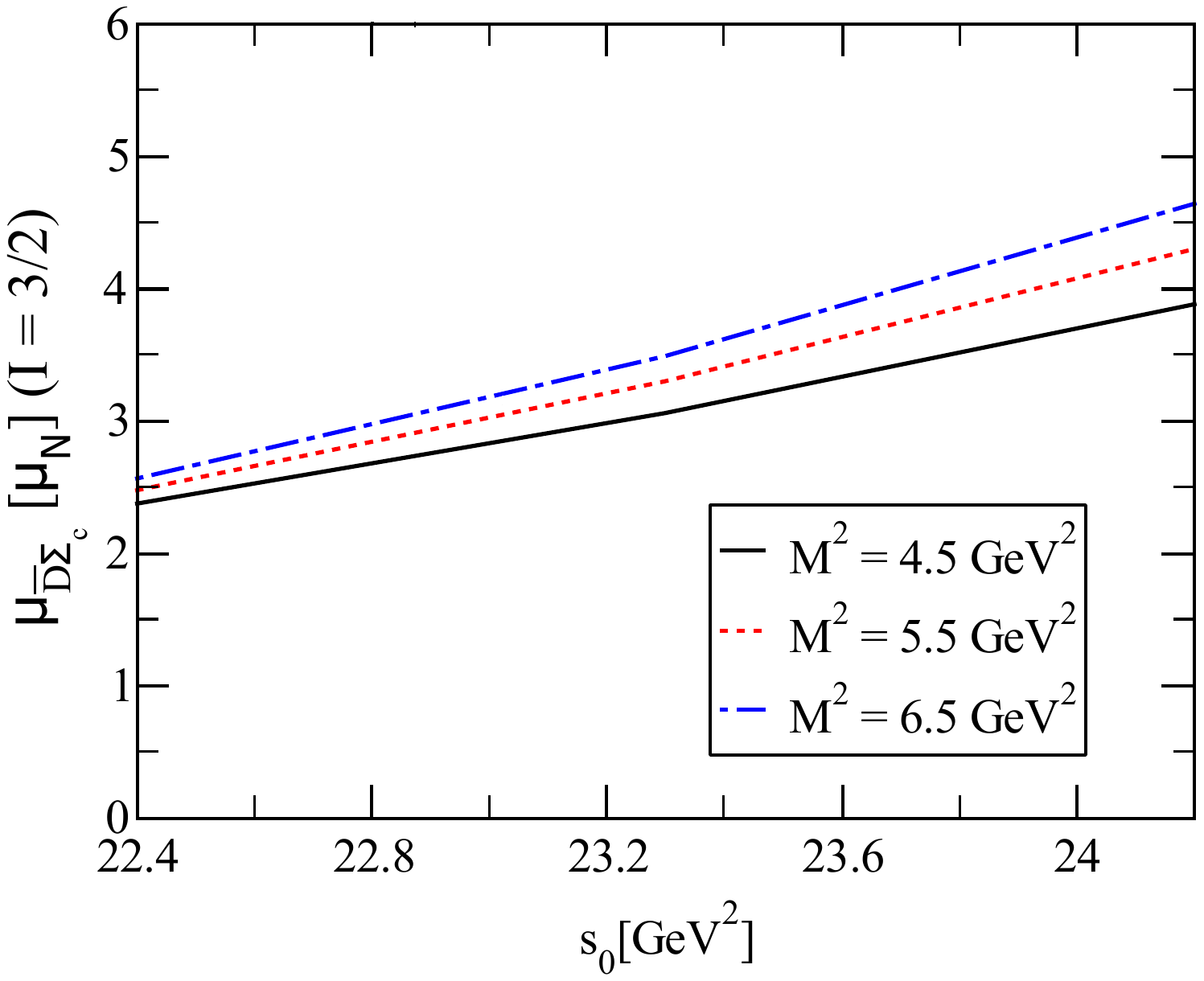}}\\
\subfloat[]{\includegraphics[width=0.4\textwidth]{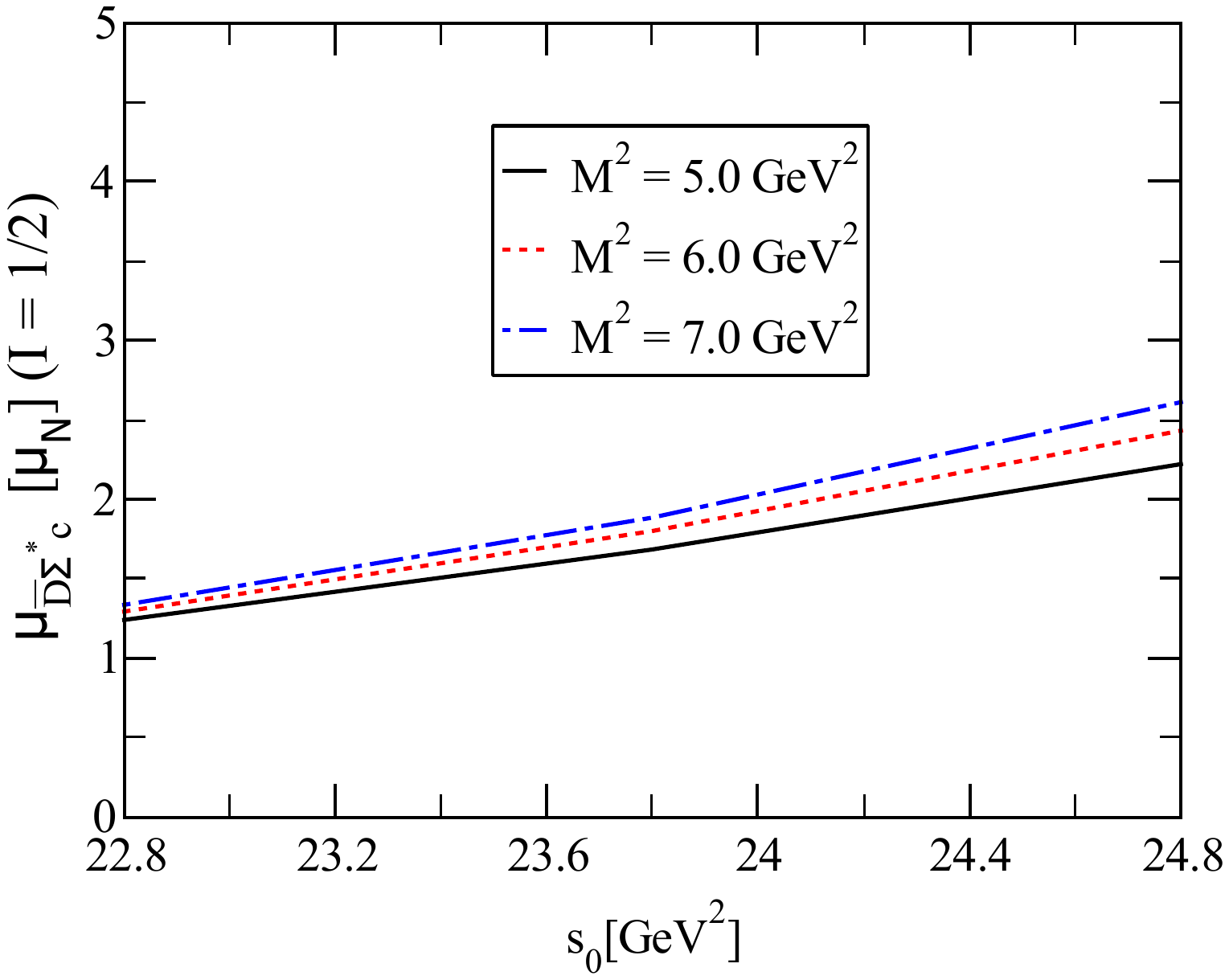}}~~~~~~~~~~~~~~~~~~
\subfloat[]{\includegraphics[width=0.4\textwidth]{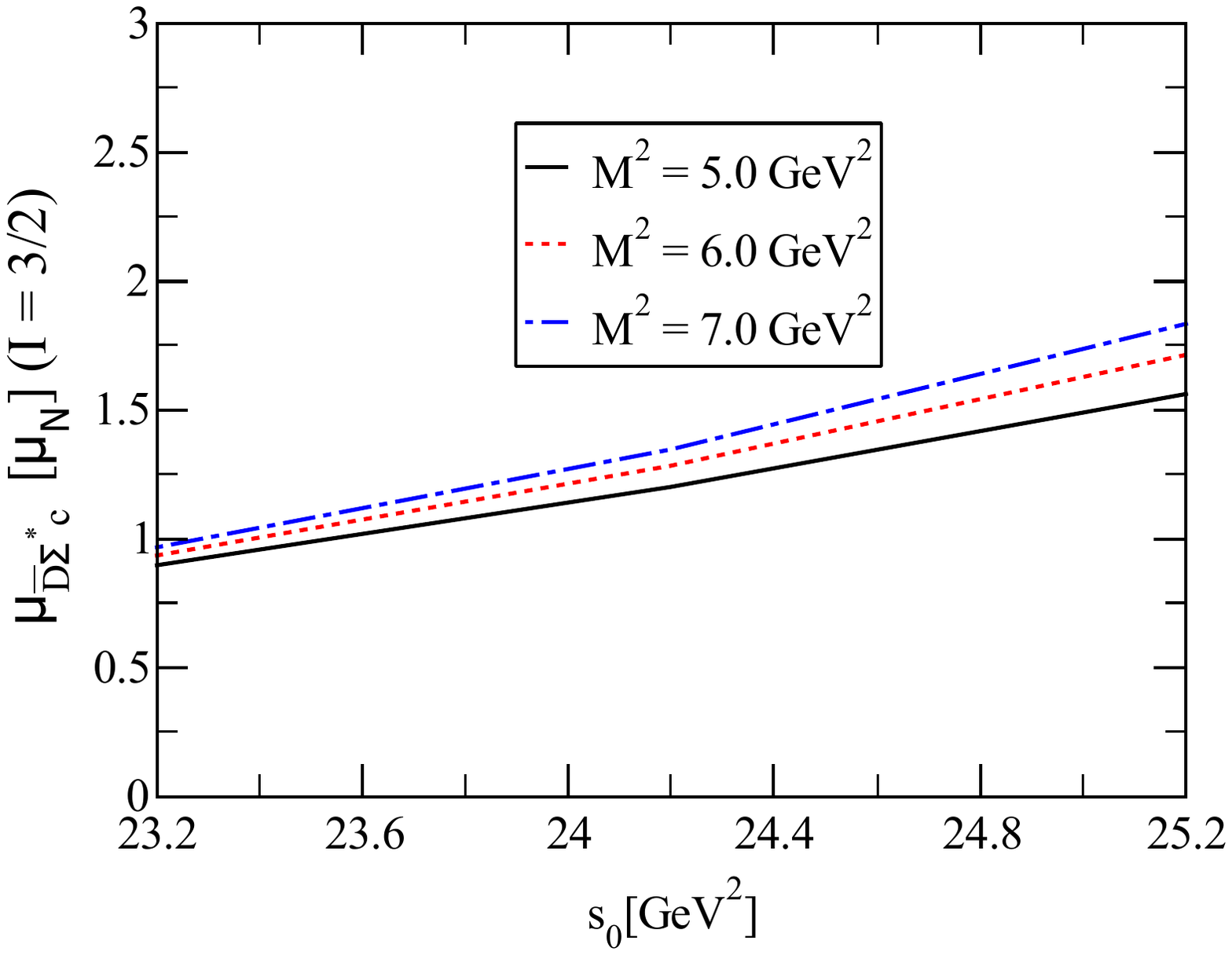}}\\
\subfloat[]{\includegraphics[width=0.4\textwidth]{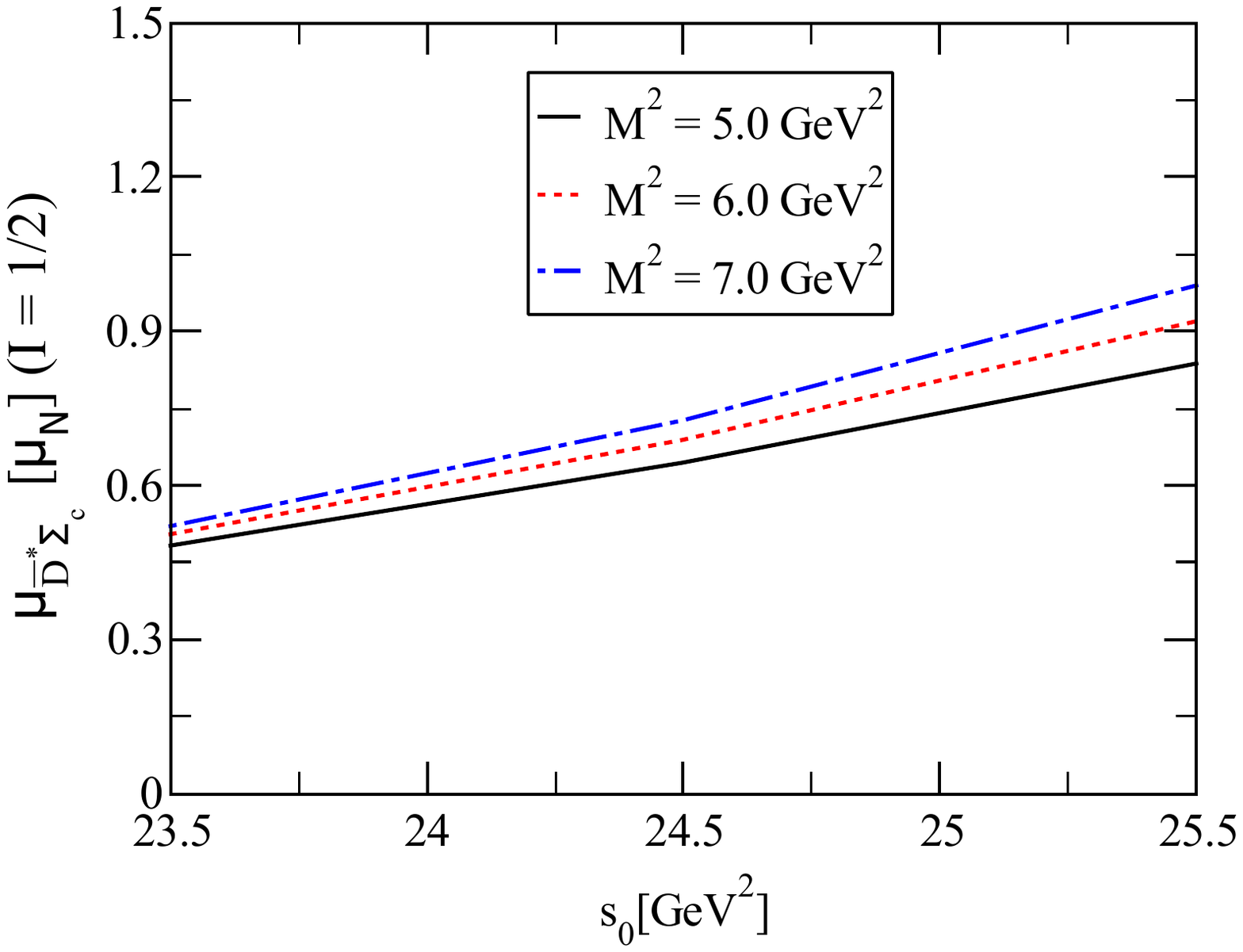}}~~~~~~~~~~~~~~~~~~
\subfloat[]{\includegraphics[width=0.4\textwidth]{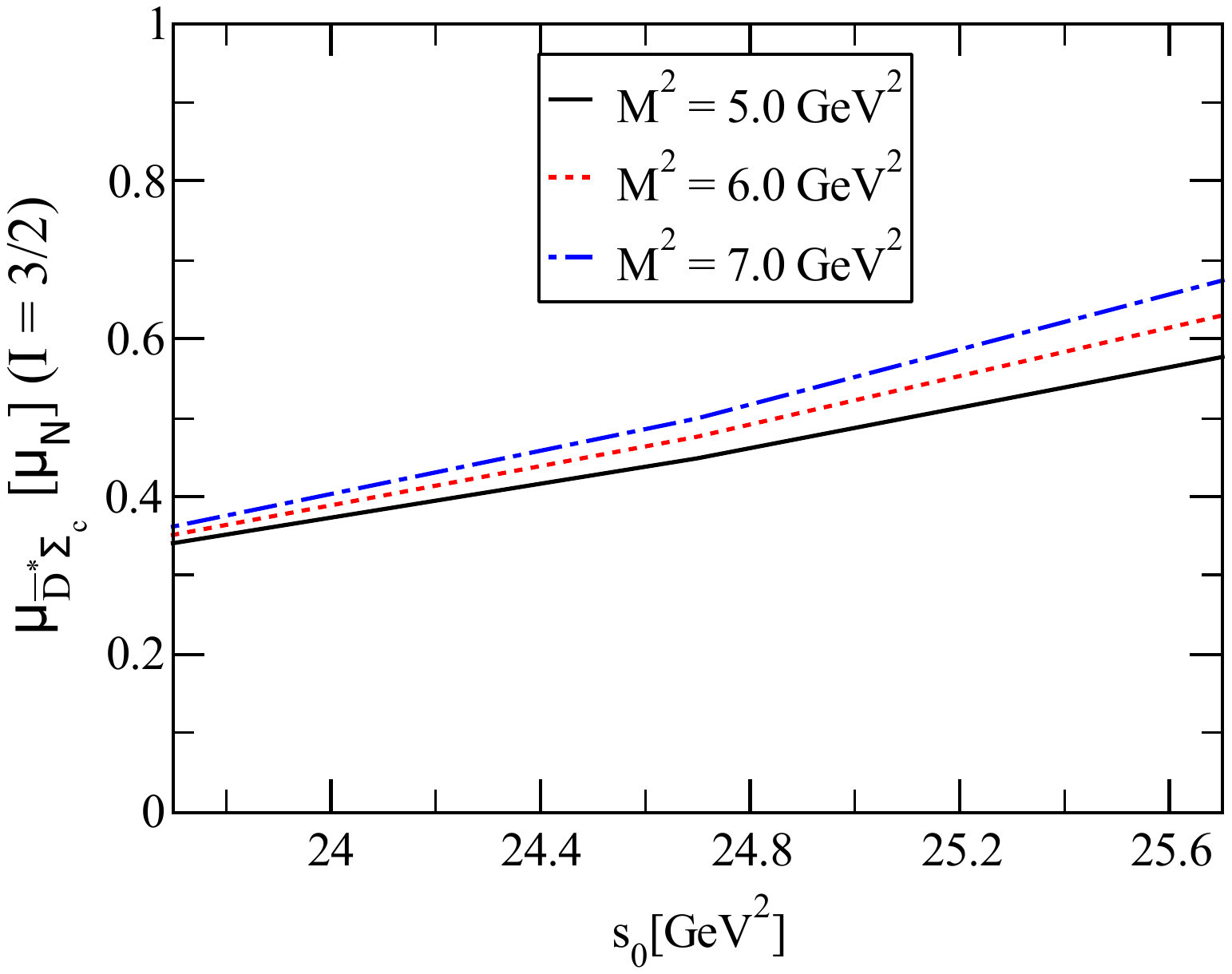}}
 \caption{The magnetic  moments of the isospin eigenstate $\bar D \Sigma_c$, $\bar D^{*} \Sigma_c$, and $\bar D \Sigma_c^{*}$ pentaquarks versus $\rm{s_0}$ at three different values of $\rm{M^2}$.}
 \label{s0fig1}
  \end{figure}

  \end{widetext}

 The continuum threshold is not completely arbitrary and is the scale where the excited states and the continuum begin to contribute to the correlation function. There are several proposed methodologies for determining this parameter. One method is to vary this parameter within a reasonable range until a Borel window emerges in which the predictions are independent of the Borel parameter~\cite{Ligeti:1993qd}. Another proposed methodology is to choose $\rm{s_0}$ as a function of the Borel parameter and to determine the functional dependence by requiring the independence of the mass prediction on the Borel parameter \cite{Lucha:2009uy}. Additionally, another method for optimizing this parameter exists, whereby the continuum parameter can be checked independently~\cite{Chen:2015moa,Chen:2016otp}.  But, the generally accepted rule for determining this parameter is to assume that $\rm{s_0} = (M_H + 0.5^{+0.1}_{-0.1}) \rm{GeV }^2$. It is recommended that the experimental data on the mass gaps between the ground states (1S) and first radial excited states (2S) pentaquarks be consulted to obtain the continuum threshold parameters $\rm{s_0}$. However, since experimental data on the excited states of hidden-charm pentaquarks is currently unavailable, a similar approach can be developed to investigate this phenomenon through possible tetraquark candidates. If we prefer the tetraquark state scenarios to other interpretations, we can assign $X(3915)$ and $X(4500)$ as the 1S and 2S hidden-charm tetraquark states, respectively, with quantum numbers $J^{PC}=0^{++}$~\cite{Lebed:2016yvr,Wang:2016gxp}, given the possible quantum numbers, decay channels, and mass discrepancies. The $Z_c(3900)$ and $Z_c(4430)$ can be identified as the 1S and 2S hidden-charm tetraquark states with the quantum numbers 
 $J^{PC}=1^{+-}$~\cite{Maiani:2014aja, Nielsen:2014mva, Wang:2014vha, Agaev:2017tzv}, respectively. The $Z_c(4020)$ and $Z_c(4600)$ can be assigned as the 1S and 2S hidden-charm tetraquark states with the quantum numbers $J^{PC}=1^{+-}$~\cite{Chen:2019osl, Wang:2019hnw}, respectively. Finally, the $X(4140)$ and $X(4685)$ were determined to be the 1S and 2S hidden-charm tetraquark states with the quantum numbers $J^{PC}=1^{++}$~\cite{Wang:2021ghk}. As can be seen, the mass discrepancies between the 1S and 2S hidden-charm tetraquark states are estimated to be within a range of $0.5-0.6 \,\rm{GeV}$. It can therefore be reasonably assumed that a comparable situation may be applicable to the excited states of hidden-charm pentaquarks. Consequently, the continuum threshold parameters can be set to $\rm{s_0} = (M_H + 0.5^{+0.1}_{-0.1}) \rm{GeV }^2$. The framework described above, in conjunction with the physical constraints defined (PC and CVG), allows for the determination of the working intervals of $\rm{M^2}$ and $\rm{s_0}$ for the purposes of this analysis.  These intervals are presented in Table  \ref{parameter} for the states that have been studied. The values of PC and CVG obtained from the analyses for each state are also presented in this table. In Fig. \ref{Msqfig1} we also show the variation of the magnetic moments of the $\bar D \Sigma_c$, $\bar D^{*} \Sigma_c$, and $\bar D \Sigma_c^{*}$ pentaquarks, on $\rm{M^2}$ for different values of $\rm{s_0}$. As can be seen from the figure, the requirement for a mild variation of the results obtained in these regions is fulfilled as required. The residual dependencies appear as uncertainties in the results. For completeness, the $\rm{s_0}$ dependence of the magnetic moments is also shown in Fig. \ref{s0fig1}. As can be seen, the deviation of the results in connection with the $\rm{s_0}$ is substantial, which is the main source of the uncertainty in the results. Note that the uncertainties in all input parameters are included in the figures.
\begin{widetext}

 \begin{table}[htb!]
	\addtolength{\tabcolsep}{10pt}
	\caption{Working intervals of  $\rm{s_0}$ and  $\rm{M^2}$ as well as the PC  and CVG for the magnetic moments of isospin eigenstate $\bar D \Sigma_c$, $\bar D^{*} \Sigma_c$, and $\bar D \Sigma_c^{*}$ pentaquarks.}
	\label{parameter}
	\begin{ruledtabular}
\begin{tabular}{l|cccccc}
               \\
State & $J^P$&Isospin&$\rm{s_0}$ (GeV$^2$)& 
$\rm{M^2}$ (GeV$^2$) & ~~  PC ($\%$) ~~ & ~~  CVG  
 ($\%$) \\
 \\
                                        \hline\hline
                                        \\
$\bar D \Sigma_c$&$ \frac{1}{2}^-$&$\frac{1}{2}$ & $22.4-24.2$ & $4.5-6.5$ & $32-58$ &  $1.82$  
                        \\
                        \\
$\bar D \Sigma_c$&$ \frac{1}{2}^-$&$\frac{3}{2}$ & $22.4-24.2$ & $4.5-6.5$ & $31-58$ &  $1.80$ 
                        \\
                        \\
$\bar D \Sigma^*$&$ \frac{3}{2}^-$&$\frac{1}{2}$ & $22.8-24.8$ & $5.0-7.0$ & $31-55$ &  $1.90$  
                       \\
                        \\
$\bar D \Sigma^*$&$ \frac{3}{2}^-$&$\frac{3}{2}$ & $23.2-25.2$ & $5.0-7.0$ & $30-55$ &  $1.80$  
                       \\
                        \\
$\bar D^* \Sigma_c$&$ \frac{3}{2}^-$&$\frac{1}{2}$ & $23.5-25.5$ & $5.0-7.0$ & $31-54$ &  $1.92$   
                      \\
                       \\
$\bar D^* \Sigma_c$&$ \frac{3}{2}^-$&$\frac{3}{2}$ & $23.7-25.7$ & $5.0-7.0$ & $30-53$ &  $1.90$   
                      \\
                       \\
 \end{tabular}
\end{ruledtabular}
\end{table}

\end{widetext}

Once all the input parameters are determined, the final predictions for the magnetic moments of the isospin eigenstate $\bar D \Sigma_c$, $\bar D^{*} \Sigma_c$, and $\bar D \Sigma_c^{*}$ pentaquarks are given in Table~\ref{sonuc}.  The errors shown in the numerical values are due to the uncertainties in the values of all input parameters, as well as the uncertainties resulting from the calculations of the working intervals for the helping parameters $\rm{M^2}$ and $\rm{s_0}$.  We would like to point out that roughly 17$\%$ of the errors in the numerical results are due to the mass of pentaquarks, 18$\%$ belongs to the residue of pentaquarks, 35$\%$ belongs to $\rm{s_0}$, 8$\%$ belongs to $\rm{M^2}$, 10$\%$ belongs to photon DAs and the remaining 12$\%$ corresponds to other input parameters. We should mention here that the QCD sum rule calculations indicate that there is one $\bar D \Sigma^*$ molecular pentaquark candidate with a mass of about $4.38$ GeV with isospin-1/2, and it does not have to be the $\rm{P_c(4380)}$. We calculate the magnetic moment of this state, and since its mass is of the value mentioned, we chose this nomenclature.

The findings of our study can be summarized as follows: 
 \begin{itemize}
  \item The orders of the magnetic moments indicate that they are accessible in the experiment. 
  
  \item Our calculations also show that the magnetic moments of  $\bar D \Sigma_c$ states with isospin-1/2 and 3/2  are governed by the light quarks, while those of $\bar D^{*} \Sigma_c$, and $\bar D \Sigma_c^{*}$ states with isospin-1/2 and 3/2 are governed by the charm quark. The discrepancy in the results may be attributed to the differing diquark components of the interpolating currents employed in the calculations, which yield disparate outcomes in the analysis. The interpolating currents constructed in different diquark structures have a profound impact on the dynamics of the system under study. For instance, while it reduces the contribution of terms proportional to the charm quark in the magnetic moment of the $\bar D \Sigma_c$ pentaquark, it can lead to the opposite situation for the $\bar D^{*} \Sigma_c$ and $\bar D \Sigma_c^{*}$ pentaquarks.  
  
  \item  A more detailed analysis shows that for the magnetic moment of the pentaquark $\rm{P_c(4312)}$, $75\%$ of the contribution comes from the $\bar D^0 \Sigma_c^+$ component.  For the pentaquark magnetic moment $\rm{P_c(4380)}$, $68\%$ of the contribution comes from the $\bar D^{0*}\Sigma_c^+$ component, while for the pentaquark $\rm{P_c(4440)}$, $60\%$ of the contribution comes from the $\bar D^0\Sigma_c^{+*}$ component.
  
  \item For the completeness of the analysis, the electric quadrupole and magnetic octupole moments of the spin-3/2 $\bar D^{*} \Sigma_c$ and $\bar D \Sigma_c^{*}$ pentaquark states are also obtained. It is well known that the sign of the electric quadrupole moment gives information about the shape of hadrons. Therefore, we can say that the shape of these pentaquarks is oblate. 
 \end{itemize}

 \begin{widetext}

 \begin{table}[htp]
	\addtolength{\tabcolsep}{10pt}
	\caption{The magnetic moments of isospin eigenstate $\bar D \Sigma_c$, $\bar D^{*} \Sigma_c$, and $\bar D \Sigma_c^{*}$ pentaquarks. We have also introduced the electric quadrupole ($\mathcal Q$) and the magnetic octupole ($\mathcal O$) moments of the $\bar D^{*} \Sigma_c$, and $\bar D \Sigma_c^{*}$ pentaquarks.}
	\label{sonuc}
		\begin{ruledtabular}
\begin{tabular}{l|cccccc}
                \\
State & $\rm{J^P}$&Isospin&$\mu\,(\mu_N)$& $\mathcal Q$\,(fm$^2$)($\times 10^{-2}$) & $\mathcal O$\,(fm$^3$)($\times 10^{-3}$) & Pentaquark \\
 \\
                                        \hline\hline
                                        \\
$\bar D \Sigma_c$&$ \frac{1}{2}^-$&$\frac{1}{2}$ & $2.59^{+0.92}_{-0.81}$ & - & -  & $\rm{P_c(4312)}$
                        \\
                        \\
$\bar D \Sigma_c$&$ \frac{1}{2}^-$&$\frac{3}{2}$ & $3.44^{+1.20}_{-1.06}$ & - & - &-
                        \\
                        \\
$\bar D \Sigma_c^*$&$ \frac{3}{2}^-$&$\frac{1}{2}$ & $1.88^{+0.73}_{-0.64}$ & $-1.92^{+0.62}_{-0.56}$ & $-0.10^{+0.04}_{-0.03}$  & $\rm{P_c(4380)}$
                       \\
                        \\
$\bar D \Sigma_c^*$&$ \frac{3}{2}^-$&$\frac{3}{2}$ & $1.34^{+0.50}_{-0.44}$ & $-1.21^{+0.38}_{-0.34}$ & $-0.06^{+0.03}_{-0.02}$  &-
                       \\
                        \\
$\bar D^* \Sigma_c$&$ \frac{3}{2}^-$&$\frac{1}{2}$ & $0.73^{+0.26}_{-0.24}$ & $-0.66^{+0.21}_{-0.20}$ & $-0.037^{+0.013}_{-0.012}$  &$\rm{P_c(4440)}$  
                      \\
                       \\
$\bar D^* \Sigma_c$&$ \frac{3}{2}^-$&$\frac{3}{2}$ & $0.50^{+0.17}_{-0.16}$ & $-0.43^{+0.12}_{-0.11}$ & $-0.023^{+0.008}_{-0.007}$&-    
                      \\
                       \\
 \end{tabular}
\end{ruledtabular}
\end{table}

\end{widetext}

To gain further insight, it would be beneficial to compare the numerical values with the results presented in the existing literature. 
In Ref. \cite{Wang:2016dzu}, the magnetic moments of the $\rm{P_c(4312)}$, $\rm{P_c(4380)}$, and $\rm{P_c(4440)}$  states are extracted within the quark model and the predicted results are $\mu_{\rm{P_c(4312)}}  = 1.76 \mu_N $, $\mu_{\rm{P_c(4380)}}  = 1.357\mu_N $, and $\mu_{\rm{P_c(4440)}}  = 3.246\mu_N$.  In Ref.~\cite{Xu:2020flp}, the magnetic moment of the $\rm{P_c(4312)}$ pentaquark state was acquired through the QCD sum rule method in the external weak electromagnetic field by using a $\bar D^{-} \Sigma_c^{++}$ molecular configuration. The numerical value was obtained as $\mu_{\rm{P_c(4312)}}  = 1.75^{+0.15}_{-0.11}$ (we would like to point out that in this study it is not written whether the magnetic moment is in its natural unit or the nuclear magneton unit). In Refs. \cite{Ozdem:2021btf, Ozdem:2018qeh, Ozdem:2021ugy}, the magnetic moments of the hidden-charm pentaquark states $\rm{P_c(4312)}$, $\rm{P_c(4380)}$, and $\rm{P_c(4440)}$   were obtained within the framework of the QCD light-cone sum rules by assuming them as $\bar D^0 \Sigma_c^+$,  $\bar D^{*-} \Sigma_c^{++}$, $\bar D^{*0} \Sigma_c^+$ molecular configurations with quantum numbers $\rm{J^P =\frac{1}{2}^-}$, $\rm{J^P =\frac{3}{2}^-}$, $\rm{J^P =\frac{3}{2}^-}$, respectively. The obtained results are presented as $\mu_{\rm{P_c(4312)}}  = 1.98 \pm 0.75\mu_N $, $\mu_{\rm{P_c(4380)}}  = 3.35 \pm 1.35\mu_N $, and $\mu_{\rm{P_c(4440)}}  = 3.49^{+1.49}_{-1.31}\mu_N$. In Ref. \cite{Li:2021ryu}, the magnetic moment of the $\rm{P_c(4312)}$ pentaquark state was extracted via quark model without/with coupled channel effects and the D-wave contributions, and the predicted result is given as  $\mu_{\rm{P_c(4312)}} = 1.624-1.737\mu_N $. In Ref. \cite{Ortiz-Pacheco:2018ccl}, the magnetic moments of four hidden-charm pentaquark states are obtained with $\rm{J^P=\frac{3}{2}^-}$ in the quark model, and the obtained results are about $(1.1-3.1)\rm{\mu_N} $. However, we should mention that all the masses of these four pentaquark states are lower than those of the observed hidden-charm pentaquark states.  From these results,  we observe that while the results obtained for the $\bar D \Sigma_c$ state are more or less similar, the results for the spin-3/2 pentaquark states $\bar D^{*}  \Sigma_c$ and $\bar D \Sigma_c^{*}$ are different in magnitude and range. 

As a final comment on comparisons, we will discuss the differences in magnetic moments obtained through the QCD light-cone sum rule method in previous studies~\cite{Ozdem:2021btf, Ozdem:2018qeh, Ozdem:2021ugy}. One important difference from previous studies on  QCD light-cone sum rules is that the states examined in this analysis are eigenstates of definite isospin. Previous studies have not considered the contribution of each component of the interpolating currents, which is examined in this analysis.  As previously stated, each component utilized in the interpolating currents has a distinct impact on the analysis, and these impacts can be substantial. Additionally, it is noteworthy that the selected interpolating currents in the analyzed studies minimize undesired contributions, such as threshold and meson-baryon scattering states, resulting in a more comprehensive and consistent analysis. This is due to the inclusion of necessary contributions and the exclusion of unwanted ones. It is generally expected that changing the basis (charge and isospin) will not affect the results. However, in the case of magnetic moments, this expectation may not hold true. This is because magnetic moments are directly related to the internal structure of the hadrons under investigation. In the language of the magnetic moment, if the basis of the hadron is changed, the internal structure of the hadron is also changed, which can substantially alter the results. In Refs.~\cite{Azizi:2023gzv, Ozdem:2024rqx, Ozdem:2022iqk}, we have used different interpolating currents, which yield the same mass values, for tetra- and pentaquarks to obtain the magnetic moment of these states, and we observe that different diquark structures yield substantially different magnetic moments. As a result, choosing different interpolating currents that potentially couple to the same states, or changing the isospin and charge basis of the states under investigation can lead to different magnetic moments. 

Before concluding the analysis, it is essential to provide a brief overview of the methodology employed to quantify the electromagnetic properties of the hidden-charm pentaquark states in the experiment. The direct measurement of the electromagnetic properties of hidden-charm pentaquark states presents a challenge due to their short lifetimes.  Consequently, the magnetic moment of this unstable state may be measured indirectly through a three-step process. The first step includes the production of the related hadron.  Subsequently, the particle emits a low-energy photon, which acts as an external magnetic field. Finally, the particle undergoes decay.    To this end, there is an alternative methodology that can be employed to ascertain the electromagnetic properties of hidden-charm pentaquark states.  The main idea of this approach is based on the hypothesis of soft photon emission proposed, which is prescribed in~\cite{Zakharov:1968fb}. The fundamental premise of this methodology is that the photon conveys information regarding the electromagnetic multipole moments of the hadron from which it is emitted. The matrix element for the radiative process can be expressed by the energy of the photon, $E_\gamma$, as

\begin{align}
 M \sim \frac{A}{E_\gamma} + B(E_\gamma )^0 + \cdots,
\end{align}
where  $\frac{1}{E_\gamma}$, $(E_\gamma)^0$, and $\cdots$  represent the contributions coming from the electric charge, the magnetic moment, and the higher multipole moments, respectively.  Consequently, the magnetic moment of the hadrons under question can be determined via the measurement of the cross-section or decay width of the radiative process, with the exclusion of the minor contributions of the linear/higher-order terms in $E_\gamma$. The magnetic moment of $\Delta$ baryon was acquired from a very similar $\gamma N $ $ \rightarrow $ $ \Delta $ $\rightarrow $ $ \Delta \gamma $ $ \rightarrow$ $ \pi N \gamma $ process~\cite{Pascalutsa:2004je, Pascalutsa:2005vq, Pascalutsa:2007wb,   Kotulla:2002cg,Drechsel:2001qu,Machavariani:1999fr,Drechsel:2000um,Chiang:2004pw,Machavariani:2005vn}.  The cross sections, whether total or differential, may be influenced by the magnetic moment of pentaquark states. A comparison can be made between theoretical prediction and experimentally determined cross-sections to ascertain the extraction magnetic moment of these states. A procedure analogous to the aforementioned methodology can be employed to investigate the electromagnetic properties of the hidden-charm pentaquarks. The production of hidden-charm pentaquarks can be achieved through either the electro-production or the photo-production process $\gamma^{(*)}N$ $\rightarrow$ $P_{c}$ $\rightarrow$ $ J /\psi N$.   Another process $\gamma^{(*)}N $ $ \rightarrow $ $P_{c} $ $\rightarrow P_{c} \gamma$ $ \rightarrow $ $ J /\psi N \gamma$ can be used to obtain the magnetic moments of the hidden-charm pentaquarks.  While it is therefore not possible at present to measure the electromagnetic properties of resonances directly experimentally, they may be derived indirectly by employing experimental data of the corresponding radiative processes.

\section{summary}\label{summary}

To shed light on the nature of the controversial and not yet fully understood exotic states, we are carrying out a systematic study of their electromagnetic properties. The magnetic moment of a hadron state is as fundamental a dynamical quantity as its mass and contains valuable information on the deep underlying structure.  In this study, we use the QCD light-cone sum rule to extract the magnetic moments of the $\mathrm{P_{c}(4312)}$, $\mathrm{P_{c}(4380)}$, and $\mathrm{P_{c}(4440)}$ pentaquarks by considering them as the molecular picture with spin-parity $\mathrm{J^P= \frac{1}{2}^-}$, $\mathrm{J^P= \frac{3}{2}^-}$, and $\mathrm{J^P= \frac{3}{2}^-}$, respectively. We define the isospin of the interpolating currents of these states, which is the key to solving the puzzle of the hidden-charm pentaquark states, to make these analyses more precise and reliable. We compared our results with other theoretical predictions, which can be a useful complementary tool for interpreting the hidden-charm pentaquark sector and we observe that they are not in mutual agreement with each other.  We have also calculated higher multipole moments for spin-3/2 $\bar D^{*}  \Sigma_c$ and $\bar D \Sigma_c^{*}$ pentaquarks, indicating a non-spherical charge distribution.

We can obtain information about the internal structure of hadrons and about the behavior of the quarks that make them up by studying their magnetic moments.  Understanding the structure and searching for it in the photo-production process would be aided by studying the magnetic moments of the pentaquarks, which could yield an independent probe of the hidden-charm pentaquarks. It will also be important to determine the branching ratios of the different decay modes and decay channels of the hidden-charm pentaquarks. Checking our predictions with future experiments could be very helpful in figuring out the geometric shape as well as the internal structure of these pentaquarks. We hope that this research, together with existing studies in the literature that include different properties of these pentaquark states, will encourage our colleagues to pay attention to the electromagnetic characteristics of pentaquark states in the future.

\begin{widetext}

  %
  \section*{Appendix: Explicit expression for \texorpdfstring{$\Delta_1^{\rm{QCD}} (\rm{M^2},\rm{s_0})$}{}}\label{appenda}
 In this appendix, we present the explicit expressions of the function $\Delta_1^{\rm{QCD}} (\rm{M^2},\rm{s_0})$ for the magnetic moment of the $\bar D \Sigma_c$ pentaquark state with isospin-1/2 entering into the sum rule.
%
\begin{align}
 \Delta_1^{\rm{QCD}}(\rm{M^2},\rm{s_0}) &=\frac {m_c} {2^{25} \times 3^3 \times5 \times 7 \pi^7} \Bigg[
   e_c \Bigg (-708 I[0, 7, 1, 3] + 2478 I[0, 7, 1, 4] - 
       3186 I[0, 7, 1, 5] + 1770 I[0, 7, 1, 6] \nonumber\\
       &- 354 I[0, 7, 1, 7] + 
       2128 I[0, 7, 2, 3] - 5319 I[0, 7, 2, 4] + 4254 I[0, 7, 2, 5] - 
       1063 I[0, 7, 2, 6] \nonumber\\
       &- 2132 I[0, 7, 3, 3] + 3196 I[0, 7, 3, 4] - 
       1064 I[0, 7, 3, 5] + 712 I[0, 7, 4, 3] - 355 I[0, 7, 4, 4] \nonumber\\
       &- 
       2478 I[1, 6, 1, 4] + 7434 I[1, 6, 1, 5] - 7434 I[1, 6, 1, 6] + 
       2478 I[1, 6, 1, 7] + 7441 I[1, 6, 2, 4] \nonumber\\
       &- 
       14882 I[1, 6, 2, 5] + 7441 I[1, 6, 2, 6] - 
       7448 I[1, 6, 3, 4] + 7448 I[1, 6, 3, 5] + 
       2485 I[1, 6, 4, 4]\Bigg) 
       \nonumber\\
       &- 
    6 (11 e_d + 106 e_u) \Bigg (3 I[0, 7, 2, 2] - 10 I[0, 7, 2, 3] + 
        12 I[0, 7, 2, 4] - 6 I[0, 7, 2, 5] + I[0, 7, 2, 6]\nonumber\\
       & - 
        9 I[0, 7, 3, 2] + 21 I[0, 7, 3, 3] - 15 I[0, 7, 3, 4] + 
        3 I[0, 7, 3, 5] + 9 I[0, 7, 4, 2] - 12 I[0, 7, 4, 3] \nonumber\\
       &+ 
        3 I[0, 7, 4, 4] - 3 I[0, 7, 5, 2] + I[0, 7, 5, 3] + 
        7 I[1, 6, 2, 3] - 21 I[1, 6, 2, 4] + 21 I[1, 6, 2, 5]\nonumber\\
       & - 
        7 I[1, 6, 2, 6] - 21 I[1, 6, 3, 3] + 42 I[1, 6, 3, 4] - 
        21 I[1, 6, 3, 5] + 21 I[1, 6, 4, 3] - 21 I[1, 6, 4, 4] \nonumber\\
       &- 
        7 I[1, 6, 5, 3]\Bigg)\Bigg]\nonumber\\
        &-\frac {m_c \langle g_s^2G^2\rangle  \langle \bar q q \rangle ^2} {2^{20} \times 3^6 \pi^3}\Bigg[ 
   3 \Bigg (-8 e_d I_ 5[\mathbb {A}] + 
      e_u \Big (5 I_ 1[\mathcal S] - 4 I_ 1[\mathcal T_ 1] + 
          4 I_ 1[\mathcal T_ 2] + I_ 1[\mathcal T_ 3] - 
          I_ 1[\mathcal T_ 4] + 5 I_ 2[\mathcal S] + 
          4 I_ 2[\mathcal T_ 1] \nonumber\\
          &- 4 I_ 2[\mathcal T_ 2] - 
          I_ 2[\mathcal T_ 3] + I_ 2[\mathcal T_ 4] + 
          6 I_ 5[\mathbb {A}]\Big)\Bigg) I[0, 2, 3, 0]
          -4 \chi \Bigg (32 e_d \Big (I[0, 3, 1, 0] - 2 I[0, 3, 1, 1] + 
       I[0, 3, 1, 2] \nonumber\\
          &- 2 I[0, 3, 2, 0] + 2 I[0, 3, 2, 1] + 
       I[0, 3, 3, 0]\Big) + 
    e_u \Big (17 I[0, 3, 1, 0] - 39 I[0, 3, 1, 1] + 
        27 I[0, 3, 1, 2] \nonumber\\
          &- 5 I[0, 3, 1, 3] - 42 I[0, 3, 2, 0] + 
        60 I[0, 3, 2, 1] - 18 I[0, 3, 2, 2] + 33 I[0, 3, 3, 0] - 
        21 I[0, 3, 3, 1] \nonumber\\
          &- 
        8 I[0, 3, 4, 0]\Big)\Bigg)\varphi_ {\gamma}[u_ 0]
        -6 \chi m_ 0^2 \Bigg ((4 e_d - 3 e_u) I_ 5[\varphi_ {\gamma}] I[0, 2, 
     3, 0] - 2 e_u \Big (I[0, 2, 1, 0] - 2 I[0, 2, 1, 1] \nonumber\\
          &+ 
       I[0, 2, 1, 2] - 2 I[0, 2, 2, 0] + 2 I[0, 2, 2, 1] + 
       I[0, 2, 3, 0]\Big) \varphi_ {\gamma}[u_ 0]\Bigg)\Bigg]\nonumber\\
       & - \frac {m_c^2 \langle g_s^2G^2\rangle  \langle \bar q q \rangle  } {2^{25} \times 3^6 \pi^5}\Bigg[  -6 e_u \Bigg (57 I_ 1[\mathcal S] + 18 I_ 1[\mathcal  T_ 1] + 
    13 I_ 1[\mathcal  T_ 2] + 12 I_ 1[\mathcal  T_ 3] - 
    43 I_ 1[\mathcal  T_ 4] + I_ 1[\mathcal  {\tilde S}] + 
    57 I_2[\mathcal S] \nonumber\\
          &- 18 I_ 2[\mathcal  T_ 1] - 
    13 I_ 2[\mathcal  T_ 2] - 12 I_ 2[\mathcal  T_ 3] + 
    43 I_ 2[\mathcal  T_ 4] + I_ 2[\mathcal  {\tilde S}] - 
    4 \Big (45 I_ 3[\mathcal S] - 29 I_ 3[\mathcal  T_ 1] + 
        29 I_ 3[\mathcal  T_ 2] - 4 I_ 3[\mathcal  T_] \nonumber\\
          &+ 
        4 I_ 3[\mathcal  T_ 4] + 32 I_ 3[\mathcal  {\tilde S}] + 
        45 I_4[\mathcal S] + 29 I_ 4[\mathcal  T_ 1] - 
        29 I_ 4[\mathcal  T_ 2] + 4 I_ 4[\mathcal  T_ 3] - 
        4 I_ 4[\mathcal  T_ 4] + 
        32 I_ 4[\mathcal  {\tilde S}]\Big)\Bigg) I[0, 3, 2, 0] \nonumber\\
          &- 
 3 \Bigg (5 (-10 e_d + e_u) I_ 1[\mathcal S] + 
    2 e_d \big (-25 I_ 1[\mathcal  {\tilde S}] + 49 I_ 3[\mathcal S] +
        I_ 3[\mathcal  {\tilde S}]\big) + 
    e_u \Big (-47 I_ 1[\mathcal  T_ 1] + 23 I_ 1[\mathcal  T_ 2] - 
        4 I_ 1[\mathcal  T_ 3] \nonumber\\
          &+ 28 I_ 1[\mathcal  T_ 4] + 
        10 I_ 1[\mathcal  {\tilde S}] + 5 I_ 2[\mathcal S] + 
        47 I_ 2[\mathcal  T_ 1] - 23 I_ 2[\mathcal  T_ 2] + 
        2 \big (2 I_ 2[\mathcal  T_ 3] - 14 I_ 2[\mathcal  T_ 4] + 
            5 I_ 2[\mathcal  {\tilde S}] + 15 I_ 3[\mathcal S]\nonumber\\
          & + 
            7 I_ 3[\mathcal  T_ 1] - 7 I_ 3[\mathcal  T_ 2] - 
            12 I_ 3[\mathcal  T_ 3] + 12 I_ 3[\mathcal  T_ 4] + 
            8 I_ 3[\mathcal  {\tilde S}] + 15 I_ 4[\mathcal S] - 
            7 I_ 4[\mathcal  T_ 1] + 7 I_ 4[\mathcal  T_ 2] + 
            12 I_ 4[\mathcal  T_ 3]  \nonumber\\
          &- 12 I_ 4[\mathcal  T_ 4]+ 
            8 I_ 4[\mathcal  {\tilde S}]\Big)\Big)\Bigg) I[0, 3, 3, 0]
            +768 f_ {3\gamma} \pi^2 \Bigg (3 e_u m_ 0^2 \Big (-I[0, 1, 1, 0] + 
       I[0, 1, 1, 1] + I[0, 1, 2, 0]\Big) 
       \nonumber\\
          &+ 
    6 e_d \Big (5 I[0, 2, 1, 0] - 8 I[0, 2, 1, 1] + 3 I[0, 2, 1, 2] - 
       8 I[0, 2, 2, 0] + 6 I[0, 2, 2, 1] + 3 I[0, 2, 3, 0]\Big) 
       \nonumber
        \end{align}
        \begin{align}
        &+ 
    2 e_u \Big (7 I[0, 2, 1, 0] - 11 I[0, 2, 1, 1] + 
        4 I[0, 2, 1, 2] - 14 I[0, 2, 2, 0] + 11 I[0, 2, 2, 1] + 
        7 I[0, 2, 3, 0]\Big)\Bigg) \psi^a[u_0]\nonumber\\
        & + 4 \chi \Bigg (6 e_d \Big (40 I[0, 4, 1, 1] - 88 I[0, 4, 1, 2] + 
        48 I[0, 4, 1, 3] - 3 I[0, 4, 2, 0] - 74 I[0, 4, 2, 1] + 
        85 I[0, 4, 2, 2] + 6 I[0, 4, 3, 0] \nonumber\\
        &+ 34 I[0, 4, 3, 1] - 
        3 I[0, 4, 4, 0]\Big) + 
     e_u \Big (216 I[0, 4, 1, 1] - 438 I[0, 4, 1, 2] + 
         260 I[0, 4, 1, 3] - 38 I[0, 4, 1, 4] + 6 I[0, 4, 2, 0]
         \nonumber\\
        &- 
         441 I[0, 4, 2, 1] + 492 I[0, 4, 2, 2] - 89 I[0, 4, 2, 3] - 
         12 I[0, 4, 3, 0] + 222 I[0, 4, 3, 1] - 48 I[0, 4, 3, 2] + 
         6 I[0, 4, 4, 0] 
         \nonumber\\
        &+ 
         3 I[0, 4, 4, 1]\Big)\Bigg) \varphi_ {\gamma}[u_ 0]                     \Bigg]\nonumber\\
         &+ \frac{\langle g_s^2G^2\rangle f_{3\gamma} m_c^2}{2^{29} \times 3^5 \pi^5}\Bigg[
         3 e_u \Big (62 I[0, 4, 3, 0] - 
    7 I[0, 4, 4, 0]\Big) I_ 2[\mathcal A] + \Big (192 e_d I[0, 4, 3, 
      0] + 54 e_u I[0, 4, 3, 0] + 624 e_d I[0, 4, 4, 0] \nonumber\\
        &+ 
    577 e_u I[0, 4, 4, 0]\Big) I_ 2[\mathcal V] + \Big (128 e_d I[0, 
      4, 3, 0] + 1398 e_u I[0, 4, 3, 0] - 146 e_d I[0, 4, 4, 0] + 
    4945 e_u I[0, 4, 4, 0]\Big) I_ 1[\mathcal V] \nonumber\\
    &+64 \Big ((3 e_d - e_u) I_ 5[\psi^a] + 
            8 e_u I_ 6[\psi_{\gamma}^{\nu}]\Big) I[0, 4, 3, 0] - 
     32 \Big ((390 e_d - 1097 e_u) I_ 5[\psi^a] + 
        16 e_u I_ 6[\psi_{\gamma}^{\nu}]\Big) I[0, 4, 4, 0]\nonumber\\
        &
        -16 \Bigg (24 e_d \Big (21 I[0, 4, 1, 0] - 79 I[0, 4, 1, 1] + 
       113 I[0, 4, 1, 2] - 73 I[0, 4, 1, 3] + 18 I[0, 4, 1, 4] - 
       77 I[0, 4, 2, 0]\nonumber\\
       &+ 216 I[0, 4, 2, 1] - 205 I[0, 4, 2, 2] + 
       66 I[0, 4, 2, 3] + 91 I[0, 4, 3, 0] - 167 I[0, 4, 3, 1] + 
       78 I[0, 4, 3, 2] - 35 I[0, 4, 4, 0] \nonumber\\
       &
       + 30 I[0, 4, 4, 1]\Big) + 
    e_u \Big (-3024 I[0, 4, 1, 0] + 8688 I[0, 4, 1, 1] - 
        7912 I[0, 4, 1, 2] + 1856 I[0, 4, 1, 3] + 392 I[0, 4, 1, 4] 
        \nonumber\\
       &+ 
        11091 I[0, 4, 2, 0] - 20765 I[0, 4, 2, 1] + 
        8241 I[0, 4, 2, 2] + 1433 I[0, 4, 2, 3] - 
        13113 I[0, 4, 3, 0] + 11434 I[0, 4, 3, 1] 
        \nonumber\\
       &+ 
        1687 I[0, 4, 3, 2] + 5049 I[0, 4, 4, 0] + 643 I[0, 4, 4, 1] - 
        3 I[0, 4, 5, 0]\Big)\Bigg)\psi^a[u_ 0]
         \Bigg]\nonumber\\
         &+\frac{m_c^2 \langle \bar qq \rangle}{2^{24} \times 3^4 \times 5 \pi^5}\Bigg[
         -12 \Bigg (2 (e_d - 2 e_u) I_ 1[\mathcal S] + 
   e_u I_ 1[\mathcal T_ 1] + 
   2 e_d \big (I_ 1[\mathcal {\tilde S}] - 3 I_3[\mathcal S] - 
      3 I_3[\mathcal {\tilde S}]\big) - 
   e_u \Big (2 I_ 1[\mathcal T_ 2] - I_ 1[\mathcal T_ 3] + 
       2 I_ 1[\mathcal {\tilde S}] \nonumber\\
       &+ 4 I_ 2[\mathcal S] + 
       I_ 2[\mathcal T_1] - 2 I_ 2[\mathcal T_ 2] + I_ 2[\mathcal T_ 3] + 
       2 I_ 2[\mathcal {\tilde S}] - 
       6 \big (3 I_ 3[\mathcal S] - 2 I_ 3[\mathcal T_1] + 
           2 I_ 3[\mathcal T_ 2] - I_ 3[\mathcal T_ 3] + 
           I_ 3[\mathcal T_ 4] + 2 I_ 3[\mathcal {\tilde S}] + 
           3 I_ 4[\mathcal S]\nonumber\\
       & + 2 I_ 4[\mathcal T_1] - 
           2 I_ 4[\mathcal T_ 2] + I_ 4[\mathcal T_ 3] - 
           I_ 4[\mathcal T_ 4] + 
           2 I_ 4[\mathcal {\tilde S}]\big)\Big)\Bigg) I[0, 5, 3, 0]
           -1920 f_ {3\gamma} \pi^2 \Bigg (e_d\Big (6 m_ 0^2 \big (I[0, 3, 2, 
           0] - 2 I[0, 3, 2, 1] \nonumber\\
       &+ I[0, 3, 2, 2] - 2 I[0, 3, 3, 0] + 
          2 I[0, 3, 3, 1] + I[0, 3, 4, 0]\big) + 7 I[0, 4, 2, 0] - 
       20 I[0, 4, 2, 1] + 19 I[0, 4, 2, 2] - 6 I[0, 4, 2, 3] 
       \nonumber\\
       &- 
       14 I[0, 4, 3, 0] + 26 I[0, 4, 3, 1] - 12 I[0, 4, 3, 2] + 
       7 I[0, 4, 4, 0] - 6 I[0, 4, 4, 1]\Big) + 
    e_u \Big (5 m_ 0^2 \big (I[0, 3, 2, 0] - 2 I[0, 3, 2, 1] 
    \nonumber\\
       &+ 
           I[0, 3, 2, 2] - 2 I[0, 3, 3, 0] + 2 I[0, 3, 3, 1] + 
           I[0, 3, 4, 0]\big) - 42 I[0, 4, 2, 0] + 79 I[0, 4, 2, 1] - 
        32 I[0, 4, 2, 2] - 5 I[0, 4, 2, 3] 
        \nonumber\\
       &+ 84 I[0, 4, 3, 0] - 
        74 I[0, 4, 3, 1] - 10 I[0, 4, 3, 2] - 42 I[0, 4, 4, 0] - 
        5 I[0, 4, 4, 1]\Big)\Bigg)\psi^a[u_ 0]
         \Bigg]\nonumber\\
         & +
 \frac {m_c\, 
   f_ {3\gamma}} {2^{26} \times 3^4 \times 5 \pi^5}\Bigg[
      4 \Big (-3 e_d I_ 1[\mathcal V] + 
          5 e_u \big (I_ 1[\mathcal A] + I_ 1[\mathcal V] - 
              I_ 2[\mathcal A] + I_ 2[\mathcal V]\big)\Big) I[0, 6, 4,
          0] - 9 \Big (-4 e_d \big (I_ 1[\mathcal V] - 
             6 I_ 2[\mathcal V]\big) \nonumber\\
             &+ 
          e_u \big (3 I_ 1[\mathcal A] + 193 I_ 1[\mathcal V] - 
              3 I_ 2[\mathcal A] + 25 I_ 2[\mathcal V]\big)\Big) I[0, 
         6, 5, 0] + 
       72 \Bigg (2 e_d \Big (13 I[0, 6, 2, 1] - 45 I[0, 6, 2, 2] + 
              57 I[0, 6, 2, 3] \nonumber\\
             &- 31 I[0, 6, 2, 4] + 6 I[0, 6, 2, 5] - 
              39 I[0, 6, 3, 1] + 96 I[0, 6, 3, 2] - 
              75 I[0, 6, 3, 3] + 18 I[0, 6, 3, 4] + 
              39 I[0, 6, 4, 1] 
              \nonumber\\
             &- 57 I[0, 6, 4, 2] + 
              18 I[0, 6, 4, 3] - 13 I[0, 6, 5, 1] + 
              6 I[0, 6, 5, 2]\Big) + 
           e_u \Big (-168 I[0, 6, 2, 1] + 493 I[0, 6, 2, 2]  \nonumber\\
             &- 
               471 I[0, 6, 2, 3]+ 135 I[0, 6, 2, 4] + 
               11 I[0, 6, 2, 5] + 504 I[0, 6, 3, 1] - 
               975 I[0, 6, 3, 2] + 438 I[0, 6, 3, 3]  \nonumber\\
             & + 
               33 I[0, 6, 3, 4]- 504 I[0, 6, 4, 1]+ 
               471 I[0, 6, 4, 2] + 33 I[0, 6, 4, 3] + 
               168 I[0, 6, 5, 1] + 11 I[0, 6, 5, 2]\Big)\Bigg) \psi^a[u_ 0]\Bigg],
        \end{align}
%
where the functions $I[n,m,l,k]$ and $I_i[\mathcal{F}]$ are written as follows:
\begin{align}
 I[n,m,l,k]&= \int_{4 m_c^2}^{\rm{s_0}} ds \int_{0}^1 dt \int_{0}^1 dw~ e^{-s/\rm{M^2}}~
 s^n\,(s-4\,m_c^2)^m\,t^l\,w^k,\nonumber\\
 I_1[\mathcal{F}]&=\int D_{\alpha_i} \int_0^1 dv~ \mathcal{F}(\alpha_{\bar q},\alpha_q,\alpha_g)
 \delta'(\alpha_ q +\bar v \alpha_g-u_0),\nonumber\\
  I_2[\mathcal{F}]&=\int D_{\alpha_i} \int_0^1 dv~ \mathcal{F}(\alpha_{\bar q},\alpha_q,\alpha_g)
 \delta'(\alpha_{\bar q}+ v \alpha_g-u_0),\nonumber\\
    I_3[\mathcal{F}]&=\int D_{\alpha_i} \int_0^1 dv~ \mathcal{F}(\alpha_{\bar q},\alpha_q,\alpha_g)
 \delta(\alpha_ q +\bar v \alpha_g-u_0),\nonumber\\
   I_4[\mathcal{F}]&=\int D_{\alpha_i} \int_0^1 dv~ \mathcal{F}(\alpha_{\bar q},\alpha_q,\alpha_g)
 \delta(\alpha_{\bar q}+ v \alpha_g-u_0),\nonumber\\
   I_5[\mathcal{F}]&=\int_0^1 du~ \mathcal{F}(u)\delta'(u-u_0),\nonumber\\
 I_6[\mathcal{F}]&=\int_0^1 du~ \mathcal{F}(u),
 \end{align}
 where $\mathcal{F}$ stands for the associated photon distribution amplitudes.

 \end{widetext}

\bibliographystyle{elsarticle-num}
\bibliography{Dbarsigmac_pentaquarksMM.bib}

\end{document}